\documentclass{aa}  

\usepackage{graphicx}
\usepackage{txfonts}
\usepackage{hyperref}
\usepackage{xcolor}

\begin{document}

   \title{Galaxy interactions in void substructures: Morphology and stellar populations of two triplets from CAVITY}
   \titlerunning{Galaxy interactions in voids}

   \author{G. M. Azevedo\inst{1,2}
          \and
          A. L. Chies-Santos\inst{2}
          \and
          R. Riffel\inst{2}
          \and
          I. Perez\inst{1}
          \and
          F. Ferrari\inst{4}
          \and \\
          R. S. de Souza\inst{5}
          \and
          M. Argudo-Fernandéz\inst{1, 3}
          \and
          B. Bidaran\inst{1}
          }

   \institute{Departamento de Física Teórica y del Cosmos, Universidad de Granada (UGR), Avenida de la Fuente Nueva S/N C.P. 18071, Granada
        \and
        Instituto de F\'isica, Universidade Federal do Rio Grande do Sul (UFRGS), Av. Bento Goncalves, 9500, Porto Alegre, RS, Brazil
        \and
        Instituto Universitario Carlos I de Física Teórica y Computacional,Universidad de Granada, 18071 Granada, Spain.
        \and
        Instituto de Matemática Estatística e Física, Universidade Federal do Rio Grande (IMEF --  FURG), Av Itália s/n,  Rio Grande, RS, Brazil 
        \and
        Centre for Astrophysics Research, University of Hertfordshire, College Lane, Hatfield, AL10~9AB, UK 
        }

   \date{Received month day, year; accepted month day, year}

  \abstract
    {Cosmic voids, vast, underdense regions of the Universe, serve as unique laboratories for studying galaxy evolution in isolation. Within these voids, galaxy triplets, rare systems of three close galaxies, provide crucial insights into how local interactions shape stellar populations and morphology in the absence of strong environmental perturbations.}
    {We investigate the spatially resolved stellar populations, morphologies, mass assembly histories, and dynamical properties of six galaxies from the Calar Alto Void Integral-field Treasury surveY (CAVITY). These galaxies reside in two void triplet systems, CAVITY5273X and VGS31.}
    {We employed integral field unit (IFU) spectroscopy combined with full spectral fitting using the \texttt{FADO} code, which simultaneously models stellar and nebular emission. We derived maps of stellar ages, metallicities, and star formation rates, applying the integrated nested Laplace approximation (\texttt{inla}) for spatial reconstruction. Morphological analysis was conducted using \texttt{morfometryka}, and emission-line diagnostics were employed to assess ionization mechanisms. Mass-assembly functions were computed to reveal the stellar mass evolution of the galaxies. We also compared stellar metallicities and stellar masses to the mass-metallicity relation (MMR) for void galaxies.}
    {The two triplets exhibit distinct evolutionary pathways. CAVITY52731 is a massive, quenched active galactic nucleus host, while young stellar populations and recent star formation dominate its companions and the three members of VGS31. The galaxies in both triplets show diverse mass assembly functions, with some building most of their stellar mass early and others exhibiting significant late-time star formation. All five star-forming galaxies present rapid mass growth in the last 2 Gyr (by a factor of $\sim 1.4 - 4$). Morphological analysis in the \textit{g} band of the DESI Legacy Survey and in the literature reveals widespread disturbances, including tidal tails, arcs, and asymmetries, indicating active dynamical evolution. Although galaxies in CAVITY5273X follow the expected MMR for voids, those in VGS31 deviate significantly, likely due to filamentary accretion and recent interactions.}
    {Our results demonstrate that even in the most underdense regions of the cosmic web, galaxies can experience dynamic evolution driven by local interactions and minor mergers. Void triplets challenge the traditional view of voids as passive environments and highlight the importance of small-scale structures in shaping galaxy properties.}
\maketitle

\section{Introduction}

Cosmic voids are vast, underdense regions that occupy most of the volume of the Universe, playing a crucial role in the formation and dynamical evolution of large-scale structures in the cosmic web \citep{bothun92,colless03, tegmark04, weygaert16, Moews2021,valles-perez21,Zhang2022,bermejo24}. These voids are enormous regions that usually span 20 - 50 $h^{-1}\text{Mpc}$ in size \citep{el-ad97,hoyle02,plionis02,hoyle&vogeley04,conroy05, sutter12}, are surrounded by denser filaments and walls of galaxies, and have very low densities of matter, down to $\lesssim20\%$ of the mean cosmic density \citep{weygaert&platen11, pan12}.

The voids originate from the underdensities of the primordial Universe \citep{plank20}, in the region where the expansion overcame the gravitational attraction of the matter. Such regions are where cosmic expansion occurs with the highest intensity, and as voids expand, matter is pulled toward their borders by gravity and squeezed between them, and sheets and filaments form the boundaries of voids \citep{weygaert&platen11}. Numerical simulations support models in which smaller-scale voids disappear into large ones \citep{dubinski93, weygaert&kampen93, colberg05,sheth04} and the density in the voids evolves to constant values in the local Universe, with a higher galaxy density profile toward their boundaries \citep{weygaert&platen11}.

Cosmological voids are exceptional laboratories for studying the evolution of galaxies with minimal influence from environmental effects, such as ram pressure stripping, mergers, tidal forces, thermal evaporation, and harassment, among others. In addition, the dark matter content in such regions is believed to be lower, and they expand faster than the mean of the cosmic web \citep{biswas10, weygaert&platen11, hamaus16}. For instance, mature voids emulate a low-density Friedmann-Lemaître-Robertson-Walker universe \citep{goldberg&vogeley04, lavaux12, hamaus14}, and dark matter halos in voids were formed later than their counterparts in overdense regions \citep{yang12, tojeiro17, wechsler18}.

The galaxy populations inside voids differ in various aspects from non-void galaxies, due to their different dynamical histories. For example, when compared to galaxies in filaments and walls, some studies find that void galaxies are, on average, younger, bluer, richer in \ion{H}{i}, less metal-enriched, and smaller, and have lower stellar masses, higher specific star formation rates (sSFRs) and later morphological types \citep{peebles01, rojas04, croton05, rojas05, hoyle12, kreckel12, beygu16, douglass18, florez21, pandey21, medrano22, rosas-guevara22, argudo24, curtis24}. However, other works indicate that void and non-void galaxies are very similar in properties, such as stellar masses, gas content, star formation rates (SFRs), chemical abundances, dark matter profiles, and metallicities \citep{szomoru96, patiri06, moorman14, liu15, douglass&vogeley17, douglass19, wegner19, dominguez-gomez22}. Thus, the ways in which void and non-void galaxies differ from each other are still poorly understood.

Recent results obtained with the Calar Alto Void Integral-field Treasury surveY (CAVITY, \citealt{cavity}), \cite{dominguez-gomez23a} show that the stellar mass assembly of void galaxies is slower than in denser environments. They find a bimodality in star formation histories (SFHs) in all environments (voids, filaments, walls, and clusters). One group formed less than 21.4\% of its stellar mass $\sim 12.5$ Gyr ago, while the other formed higher fractions of its stellar mass at the same time. For both cases, the SFHs in voids are the slowest. Also, \cite{dominguez-gomez23b} find that galaxies in voids exhibit slightly lower central stellar metallicities ($\sim$0.1 dex) than those in filaments and walls, and much lower ($\sim$0.4 dex) than those in clusters. These differences are more pronounced in low-mass galaxies ($\lesssim 10^{9.25} \text{M}_{\odot}$), galaxies with long-timescale SFHs, and spiral and blue galaxies. Additionally, \cite{conrado24} find that void galaxies have a slightly higher half-light radius (HLR), a lower stellar mass surface density, younger ages across all morphological types, and a slightly elevated SFR and sSFR (only significant enough for Sas). Their analysis also indicates that void galaxies undergo a more gradual evolution, especially in their outer regions, with a more pronounced effect for low-mass galaxies.

Voids are dynamic and evolving systems \citep{courtois23}. They asymptotically evolve to empty and structureless regions as the space expands and the gravitational force of the clusters and surrounding filaments and walls pull the galaxies and dark matter halos toward the void outskirts \citep{weygaert&platen11}. However, in the local Universe, the voids are still not devoid of internal structure. As those voids expand and merge hierarchically, structures such as diffuse filaments and groups of galaxies may remain inside those voids \citep{szomoru96,el-ad97,hoyle&vogeley04,kreckel12,beygu13,alpaslan14}. Some of the remaining substructures we find in voids are galaxy triplets, which are the simplest type of galaxy groups formed by three galaxies \citep{karachentseva79, karachentseva&karachentseva00, elyiv09, makarov&karachentsev09}. Such systems located in voids are very scarce, and there is no vast literature on them. Isolated galaxy triplets represent only 3\% of galaxies with magnitudes in the r band of $11 \leq m_r \leq 15.7$ at low redshifts $0.005 \leq z \leq 0.08$ \citep{argudo-fernandez15}. Also, the majority of triplets are located in the outer parts of filaments, walls, and clusters, instead of voids. Some important catalogs of triplets in the literature are \cite{karachentseva79}, \cite{trofimov&chernin95}, \cite{o'mill12}, and \cite{argudo-fernandez15}.

Most galaxy triplets exhibit signs of a long dynamical evolution whereby the system is embedded in a common dark matter halo, and the member galaxies present similar properties \citep{chernin00, hernandez-toledo11, duplancic13, feng16, emel'yanov16}. In that sense, they are much closer to compact groups (CGs; \citealt{duplancic15,costa-duarte16}), which are formed by four galaxies or more. \cite{duplancic13} found that galaxies in triplets have colors, SFRs, and stellar populations similar to CGs. Also regarding stellar populations, \cite{vasquez-bustos23} find that there is no dominant type of galaxy in isolated triplets in terms of global colors and the SFR.

An additional interest in void galaxies is that we can properly investigate the local environmental influence of the neighboring galaxies on each other and discard the interference of large-scale structures (clusters and groups) dynamics. \cite{argudo-fernandez15} found no difference in the interaction of large-scale environments with isolated galaxies, isolated pairs, and isolated triplets. This suggests that both have common origins and the differences in observational properties are due to differences in their local environments and dynamics.

This work characterizes the stellar populations of six galaxies belonging to two void triplet systems. We performed spatially resolved stellar population synthesis (SPS) on the integral field unit (IFU) data obtained by CAVITY. To perform the synthesis, we used the code {\sc fado} \citep{fado} and the front-end software \href{https://github.com/ndmallmann/urutau}{{\sc urutau}}.

The structure of this paper is organized as follows. Section 2 explains the data and methodologies used in this work. In Sect. 3 we present our results. In Sect. 4 we discuss their implications, followed by our conclusions in Sect. 5. Throughout this work we use the following cosmology: $H_0 = 67\,\text{km}\,\text{s}^{-1}\text{Mpc}^{-1}$, $\Omega_M = 0.3$, and $\Omega_{\Lambda} = 0.7$ \citep{planck19}.

\section{Data}

\subsection{CAVITY spectroscopy}

The data used in this work were obtained from CAVITY \citep{cavity}. It is a legacy survey conducted at the Calar Alto Observatory, aiming to obtain spatially resolved spectroscopy of approximately 300 void galaxies in the local Universe. These galaxies are distributed across 15 cosmological voids, with redshifts in the range of $0.005 < z < 0.05$, and have absolute r-band magnitudes between $-21.5$ and $-17.0$. The first data release was done with 100 galaxies \citep{cavityDR1}, though none of the objects studied in this work are in DR1.

Observations were carried out using the integral field Potsdam Multi-Aperture Spectrograph (PMAS; \citealt{pmas}), mounted on the 3.5m telescope at the Hispanic Astronomical Center in Andalusia (CAHA), operating in PPaK mode \citep{ppak}. PPaK consists of over 300 fibers, each with a diameter of 2.68 arcseconds, arranged to cover a hexagonal field of view (FoV) of 74 × 64 arcsec$^2$ with a filling factor of approximately 60\%. The V500 grating was used, offering a spectral range of 3475–7300 Å and a spectral resolution of $\lambda/\Delta\lambda = 850$ at 5000 Å, corresponding to a full width at half maximum (FWHM) of roughly 6 Å. The cubes were reduced and calibrated through the standard CAVITY data reduction pipeline and they were corrected for galactic extinction during this procedure (for more details, see \citealt{garcia-benito24}).

\subsection{Imaging}

In Sect. \ref{morphology} we use broad band images in order to perform the morphological study of our targets. We use the images of the $g$ band from the Dark Energy Spectroscopic Instrument (DESI) Legacy Survey \citep{dey19}. This survey uses as its main instrument the Dark Energy Camera (DECam) in the Blanco Telescope in Chile, in addition to the Mosaic3 Camera in the Mayall Telescope in EUA. Both instruments have pixels with an angular size of 0.262 arcsec, and an angular resolution between $\sim1$ and $\sim1.5$ arcsec, depending on the band. The $g$ band, by being the bluer filter in the $gri$ system, provides some of the best evidence of the asymmetries in the young populations in the galaxies.

\subsection{Sample}

\begin{figure*}
    \centering
    \includegraphics[width=0.65\linewidth]{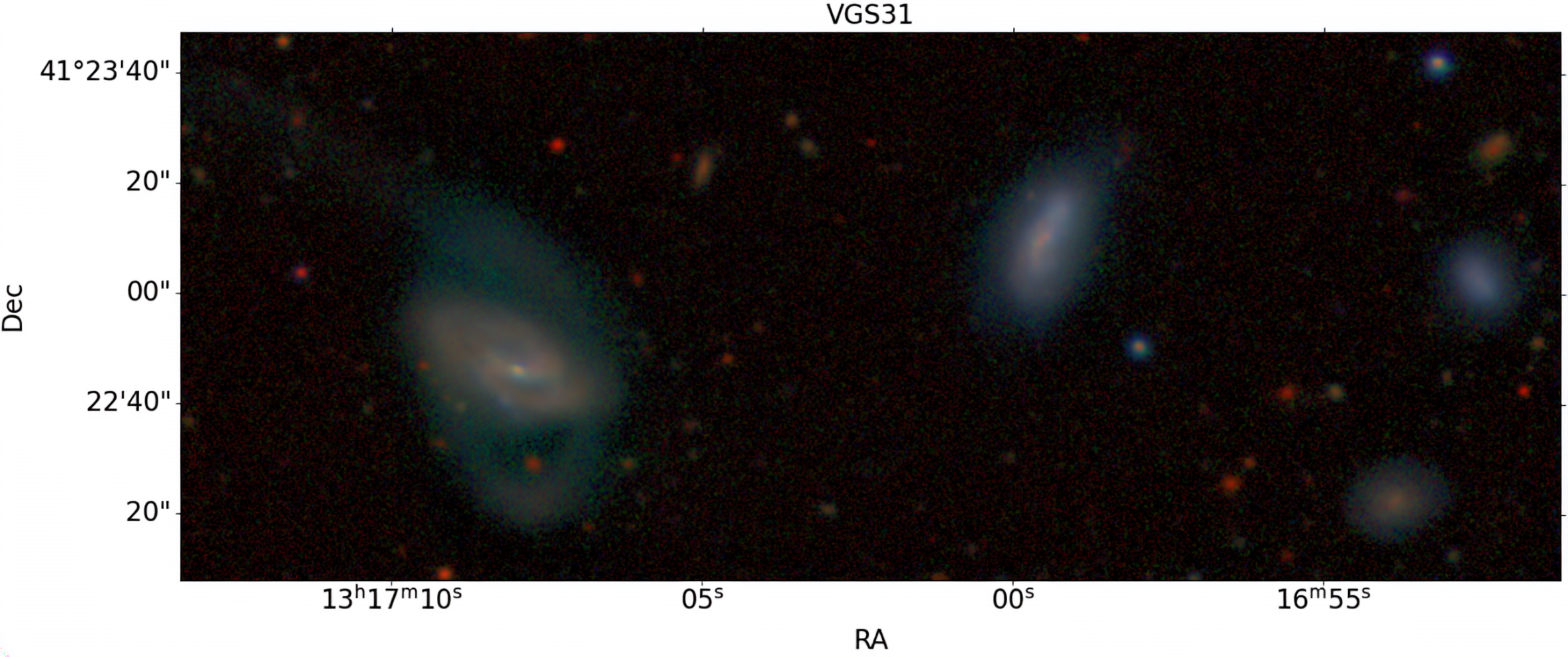} \includegraphics[width=0.335\linewidth]{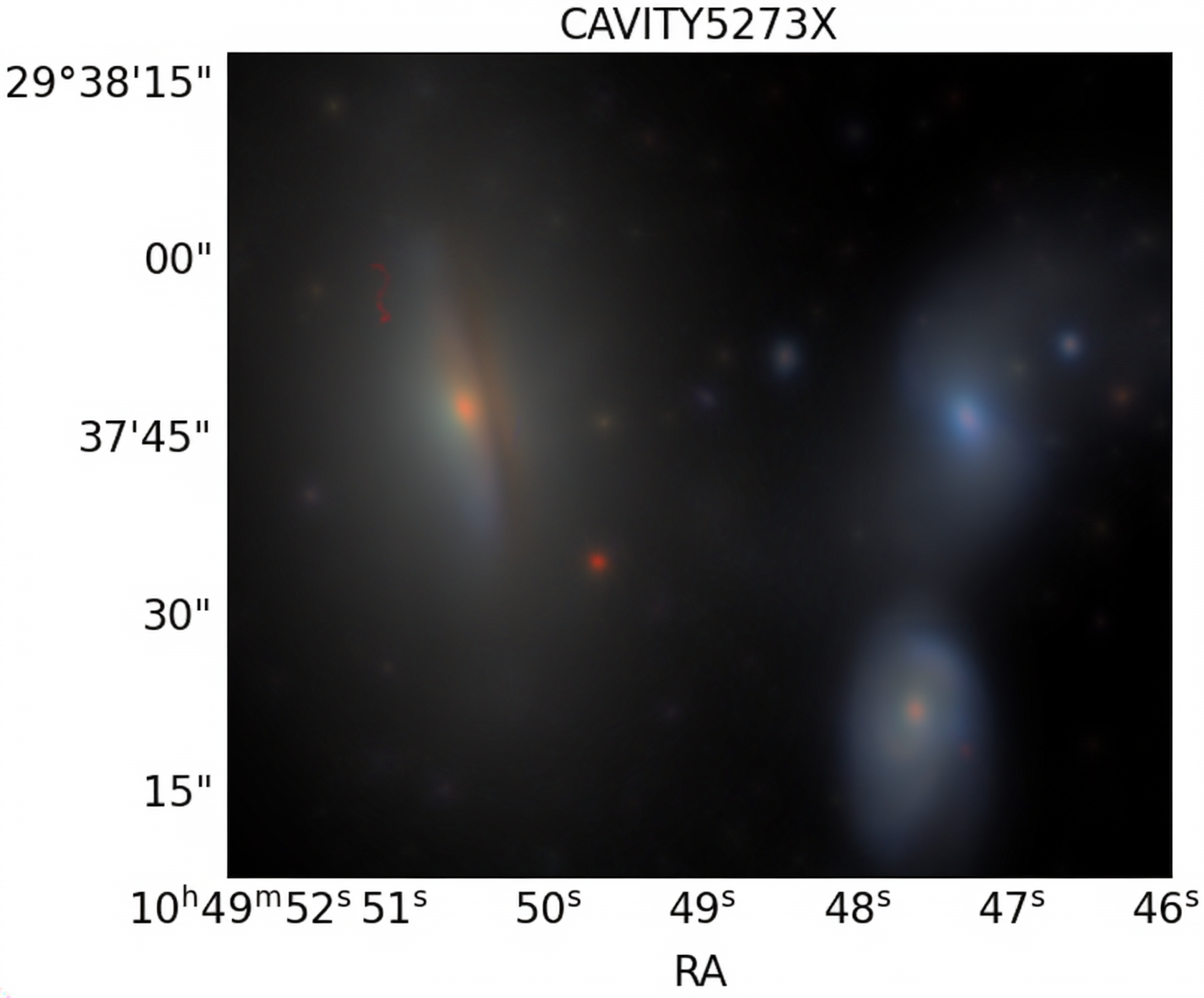}
    \caption{RGB images of the two triplets taken from DESI Legacy Survey. The left panel shows the VGS31 triplet, with VGS31a in the center, VGS31b to its southeast, and VGS31c to its west. The right panel shows CAVITY5273X, with CAVITY52731 to the east, CAVITY52732 to its west, and CAVITY52733 to its southwest.}
    \label{triplets_rgb}
\end{figure*}

The subjects of this study are two galaxy triplet systems located in distinct cosmological voids. They were the first triplets observed by the survey. Figure \ref{triplets_rgb} shows RGB images of the triplets obtained from Pan-STARRS \citep{chambers16}. We refer to these systems as CAVITY5273X and VGS31. These two specific systems were selected because they were the only triplets with all three galaxies observed by CAVITY's IFU.

There is no detailed literature available on CAVITY5273X. This system comprises a larger, redder galaxy (i.e., CAVITY52731) accompanied by two smaller, bluer companions to its west (i.e., CAVITY52732) and southwest (i.e., CAVITY52732). All three galaxies exhibit some degree of morphological asymmetry in the light, as is discussed further in Sect. \ref{morphology}. For example, CAVITY52732 features two tidal tails.

VGS31, on the other hand, is also part of the Void Galaxy Survey (VGS, \citealt{weygaert11}) and has been studied previously by \cite{beygu13}, who found it to be embedded in an \ion{H}{i} filament. This makes VGS31 an example of a filamentary structure within a void, suggesting that its constituent galaxies may be undergoing active growth. The system is composed of a central galaxy (i.e., VGS31a), a similarly sized galaxy to its east (i.e., VGS31b), and a smaller one to its west (i.e., VGS31c). All galaxies in the system show asymmetric features in the images, particularly VGS31b, which exhibits a tidal tail to the north and a ring-like arc to the south -- likely the result of a minor merger \citep{beygu13}. Table \ref{tabela1} summarizes basic previously known properties of these galaxies.

\begin{table*}[]
\centering
\caption{Previously measured basic information about the galaxies.}
\begin{tabular}{llllllllll}
\hline
Galaxy      & RA (J2000)  & Dec (J2000)  & z     & u(SDSS) & g(SDSS) & r(SDSS) & i(SDSS) & z(SDSS) \\ \hline
VGS\_31a    & 13:16:06.19 & +41:30:04.25 & 0.021 & 15.936  & 15.059  & 14.731  & 14.525  & 14.358 \\
VGS\_31b    & 13:16:14.69 & +41:29:40.05 & 0.021 & 15.886  & 14.972  & 14.457  & 14.169  & 13.965  \\
VGS\_31c    & 13:15:59.18 & +41:29:55.96 & 0.021 & 17.843  & 17.006  & 16.771  & 16.641  & 16.594  \\
CAVITY52731 & 10:49:10.05 & +29:48:57.36 & 0.031 & 16.846  & 14.857  & 14.020  & 13.566  & 13.249  \\
CAVITY52732 & 10:49:06.78 & +29:48:56.54 & 0.031 & 17.834  & 16.487  & 15.967  & 15.666  & 15.736  \\
CAVITY52733 & 10:49:07.13 & +29:48:31.46 & 0.031 & 18.101  & 16.476  & 15.952  & 15.571  & 15.338  \\ \hline
\end{tabular}
\label{tabela1}
\end{table*}

\section{Methodology}

\subsection{Spectral fitting}
\label{sps}

In order to implement the SPS in the CAVITY datacubes, we used the nonparametric code \href{http://spectralsynthesis.org/fado.html}{{\sc fado}} \citep[Fitting Analysis using Differential evolution Optimisation;][]{fado}. To achieve the best-fitting solution, {\sc fado} employs a genetic differential evolution optimization (DEO) algorithm in which each candidate model -- defined by stellar population fractions, extinction, and kinematics -- is treated as an individual within a population that evolves through mutation and recombination. At each generation, the fittest models are selected, ensuring convergence toward solutions that simultaneously reproduce the stellar continuum and the nebular emission.

The additional constraint of the nebular emission was computed from the fluxes of various emission lines (H$\alpha$, H$\beta$, [\ion{N}{ii}], [\ion{S}{ii}], [\ion{O}{iii}], etc). This constraint is critical for the synthesis of star-forming galaxies \citep{fado2}, since it enables the computation of nebular spectra in addition to the stellar ones. The software computes the number of ionizing photons required to produce the observed fluxes of the main emission lines, especially the Balmer lines, as hydrogen is ionized by photons with wavelengths shorter than 912 \AA, which are emitted by hot, massive young stars (t < 20 Myr). This way, it estimates the necessary fraction of young stellar populations to account for the observed ionization.

To run a SPS code, it is necessary to supply a base of simple stellar populations (SSPs). In this work, we use a base comprising 68 SSPs from \citet[hereafter BC03]{BC03}, with a \cite{chabrier} initial mass function (IMF), and lower and upper stellar mass limits of 0.1 $M_{\odot}$ and 100 $M_{\odot}$, respectively. Our choice of BC03 is due to its ability to consistently extend into the ionizing UV domain ($\lambda < 912$ \AA), which is required by \textsc{fado}’s self-consistent modeling of stellar and nebular emission. Among the available models, there is only one that covers the ionizing UV emission and has a spectral resolution in the optical comparable to the data in the optical: the BC03 ($\text{FWHM} \approx 3$ \AA). The BC03 models also have their limitations, particularly for the intermediate-age range, which gets under-sampled as a result of insufficient TP-AGB modelling \citep{riffel15, lu25}. However, such subrepresentation affects the NIR spectral range more than the optical. The models from \cite{M05} also cover the ionizing UV domain, but have a poorer spectral resolution in the optical ($\text{FWHM} \approx 10-20$ \AA). Although newer versions of these models present higher resolutions, such as \cite{M11}, they do not cover the ionizing domain, reaching down to 1000 \AA.

However, it is unnecessary to use the entire SSP library from BC03, since many models are redundant for synthesis purposes. Simple stellar populations with similar ages and metallicities often have nearly indistinguishable spectra, particularly for older populations (t > 5 Gyr), and the inclusion of observational noise further reduces their discriminability. Moreover, using an excessively large base greatly increases computational time without significant improvements in fitting accuracy. Instead, it is best to select a representative subset covering different age ranges. We followed the methodology described in \cite{dametto14} and also applied in \cite{azevedo23} for BC03, selecting a base of 68 SSPs with 17 representative ages (t = 1.00, 2.09, 3.02, 5.01, 10.00, 25.12, 50.00, 101.50, 321.00, 508.80, 718.70 Myr and 1.02, 2.00, 3.00, 5.00, 10.00, 13.00 Gyr) and four metallicities (Z = $\frac{1}{200}Z_{\odot}, \frac{1}{5}Z_{\odot}, 1Z_{\odot}, \frac{5}{2}Z_{\odot}$), which have seven representative ages for young populations (t < 100 Myr), seven representative ages for intermediate populations (100 Myr < t < 5 Gyr), and three for old populations (t $\geq$ 5 Gyr).

A challenge when applying SPS to datacubes is that the synthesis codes are designed for single spectra. Therefore, one must extract the spectrum from each spaxel, apply the synthesis individually, and then store the results, indexing them back to their corresponding spaxels. Fortunately, there is a front-end pipeline called {\sc urutau} \footnote{The code is available
at: \href{https://github.com/ndmallmann/urutau}{https://github.com/ndmallmann/urutau}} \citep{riffel23} that automates all these steps and, additionally, writes the synthesis results into new extensions of the original datacube. {\sc urutau} operates through modules dedicated to each processing step, such as extraction, dereddening, synthesis, and result reading, as well as offering extra features. However, the original modules were developed specifically for use with {\sc starlight} \citep{starlight}, which functions quite differently from {\sc fado}. Thus, we developed new modules to run {\sc fado} on the spectra and to parse its outputs accordingly.

\subsection{Spatial reconstruction}
When we implement SPS methodology on IFU data, the SPS needs to be applied to each spaxel independently. This can generate abrupt variations in the values of a given measurement of two neighboring pixels. For instance, a pixel can have a low mean age and its neighbor can have a high mean age, which is not physically accurate. Smoother gradients are expected for this kind of extended source \citep{inla}.

In order to avoid such cases and produce smoother morphological maps with a mathematically robust methodology, we made use of the integrated nested Laplace approximation (\href{https://www.r-inla.org/}{\sc inla}). It is an alternative method to Markov chain Monte Carlo for fast Bayesian inference. Its mathematical foundations and applications to IFU data from the CALIFA and PISCO surveys are presented in \cite{inla}. \cite{azevedo23} also applied the method to spatially resolved data from the MUSE spectrograph. {\sc inla} is highly efficient at both reconstructing data from extended sources with missing pixels and smoothing data by considering spatial correlations. Large variations between neighboring pixels for physical quantities such as stellar ages or metallicities may be unphysical. In this context, the method accounts for spatial correlations between adjacent pixels to reconstruct a more physically consistent map. There is a comprehensive R library for implementing {\sc inla}.

\begin{figure}
    \centering
    \includegraphics[width=\linewidth]{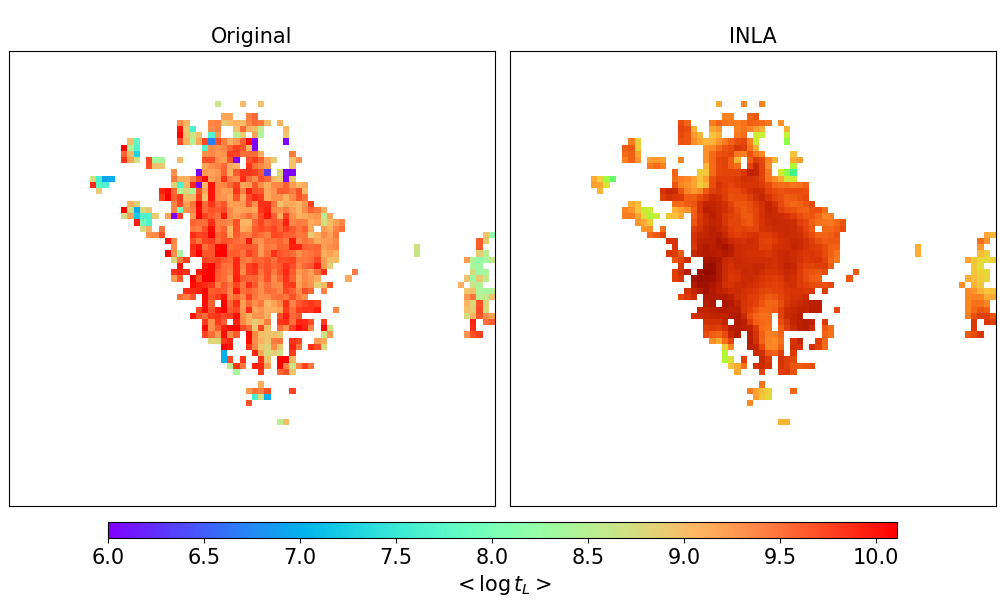}
    \caption{Comparison of the $<\log t>_L$ map for the galaxy CAVITY52731 before (left) and after (right) applying INLA.}
    \label{inla}
\end{figure}

In this work, we used it on 2D maps, such as those shown in Sect. \ref{outputs}. It also acts as a way to correlate the results of the SPS analysis from individual spaxels, which are treated independently in our synthesis, but in reality are part of the same extended source. Figure \ref{inla} presents an example of $<\log t>_L$ maps before and after applying INLA to the map.

\subsection{Signal-to-noise ratio}

\begin{figure*}
    \centering
    \includegraphics[width=\linewidth]{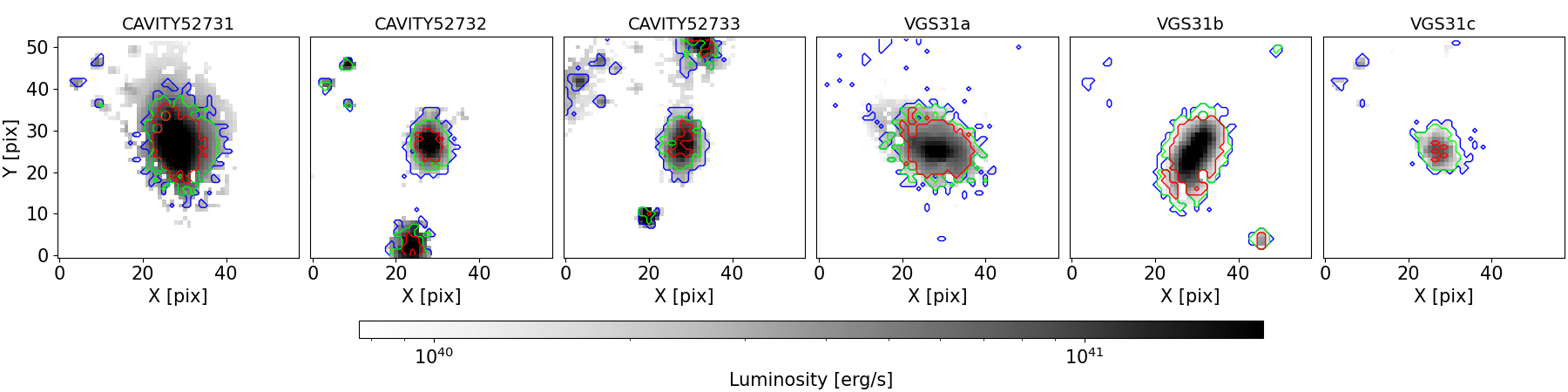}
    \caption{Maps of luminosities of the galaxies, with the colored lines indicating different contours of S/N. The red contour is for S/N$\geq10$, the green for 5, and the blue for 3. The S/Ns were computed in the wavelength interval of 4000 - 4060 \AA.}
    \label{S/N}
\end{figure*}

The reliability of a SPS depends on the signal-to-noise ratio (S/N) of the spectra. Figure \ref{S/N} shows the S/N contours overlaid on the luminosity maps of each galaxy. The S/N values were computed in the featureless continuum wavelength window from 4000 to 4060~\AA. This wavelength range is the same as that used to normalize the spectra during the SPS process, and it was chosen to maintain consistency. 

It is also close to the lower wavelength limit of our observed data. The blue end of the spectrum is significantly noisier than the red end in our data, with uncertainties in the 4000 -- 4060~\AA{} range reaching values up to ten times higher than those at the opposite end. However, this region is also the most relevant for SPS, since the slope of the NUV and blue continuum contains the most prominent differences between different stellar populations, especially the youngest ones.

Therefore, the most reliable synthesis results correspond to spaxels with $\text{S/N} \geq 10$, which generally excludes the outer regions of our galaxies. All the spaxels that appear in Fig. \ref{S/N} were used in the analysis, but it is important to keep in mind that the results on the outskirts of the galaxies are less reliable.

\subsection{Spectral fits outputs}
\label{outputs}

A SPS code provides a series of different quantities for each spectrum, both directly or that can be easily obtained from its results, such as mean stellar ages and metallicities, stellar masses, stellar and nebular radial velocities and velocity dispersions, SFRs, stellar and nebular extinctions, among many others \citep{starlight, fado}. {\sc fado}, for instance, by dealing with nebular emission physics, also returns fluxes, equivalent widths, rotation velocity, and velocity dispersion for various emission lines, in addition to the number of ionizing photons for hydrogen and helium.

The mean stellar ages are defined in two ways \citep{starlight,fado}. The mean age weighted by light can be expressed as
\begin{equation}
    \langle \log t\rangle_L = \sum_{j=1}^{N_{\star}} x_j \log(t_j)\,
.\end{equation}
Meanwhile, the mean age weighted by mass is expressed as
\begin{equation}
    \langle \log t \rangle_M = \sum_{j=1}^{N_{\star}} \mu_j \log(t_j)\,,
\end{equation}
where $t_j$ is the age of the j-th SSP from the base measured in years, $x_j$ is the fraction with which it contributes to light, $\mu_j$ is the fraction with which it contributes to mass, and $N_{\star}$ is the number of SSPs in the base. Similarly, the mean stellar metallicities are defined as
\begin{equation}
    \langle Z \rangle_L = \sum_{j=1}^{N_{\star}} x_j Z_j\,,
\end{equation}
\begin{equation}
    \langle Z \rangle_M = \sum_{j=1}^{N_{\star}} \mu_j Z_j\,,
\end{equation}
where $Z_j$ is the metallicity of the j-th SSP. The fraction of light or mass represented by each SSP is provided in what we call population vectors.

Another important measurement we can make using the population vectors is the SFR. We can compute it through the equation \citep[][and references therein]{riffel21}
\begin{equation}
    {\rm SFR}(t) = \frac{dM_{\star}^c}{dt} \approx \frac{\Delta M_{\star}^c}{\Delta t} = \frac{\sum_{j_i}^{j_f}M_{\star,j}^c}{\Delta t},
\end{equation}
\label{SFR_stars}
where $j$ refers to the $j$-th SSP in the base, and $M_{\star}^c$ is the total mass converted to stars in the time interval $\Delta t$. It is worth highlighting that the stellar mass converted to stars is different than the mass present available in stars. The present stellar mass takes into account the mass expelled by the stars during their evolution, while the other represents all the mass collapsed into stars in a certain time epoch. {\sc fado} computes the total $M_{\star}^c$ and we can then use the coefficients, $\mu_j$, to find the mass converted into stars in a given time.

\begin{figure*}
    \centering
    \includegraphics[width=\linewidth]{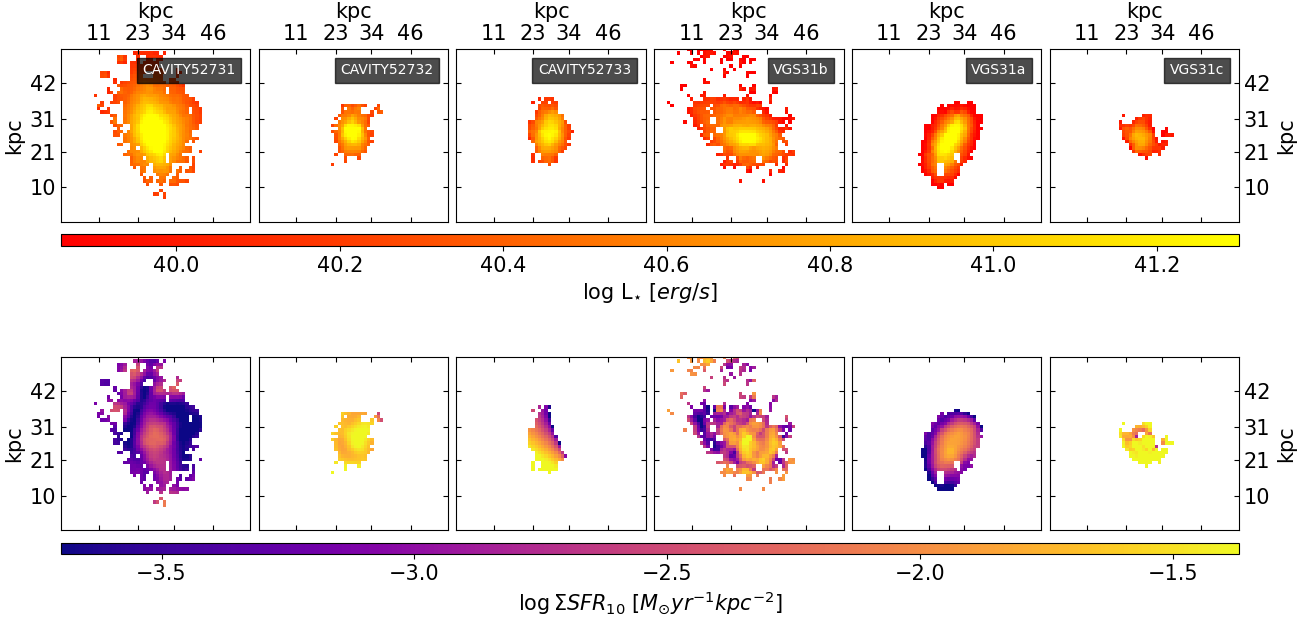}
    \includegraphics[width=\linewidth]{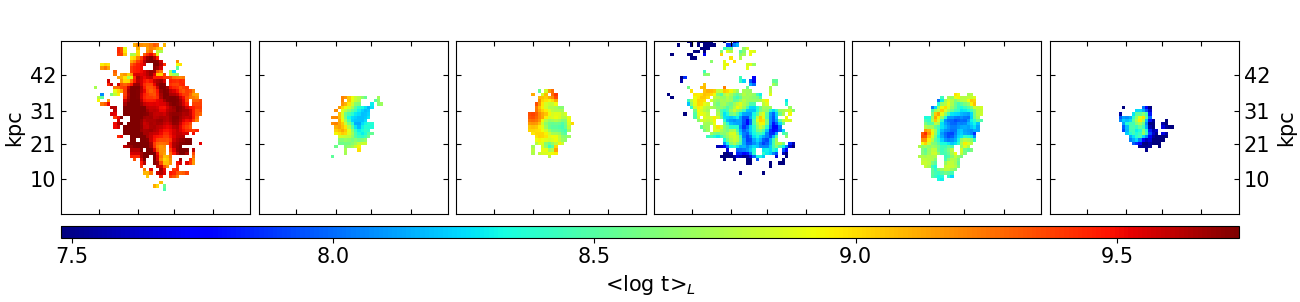}
    \includegraphics[width=\linewidth]{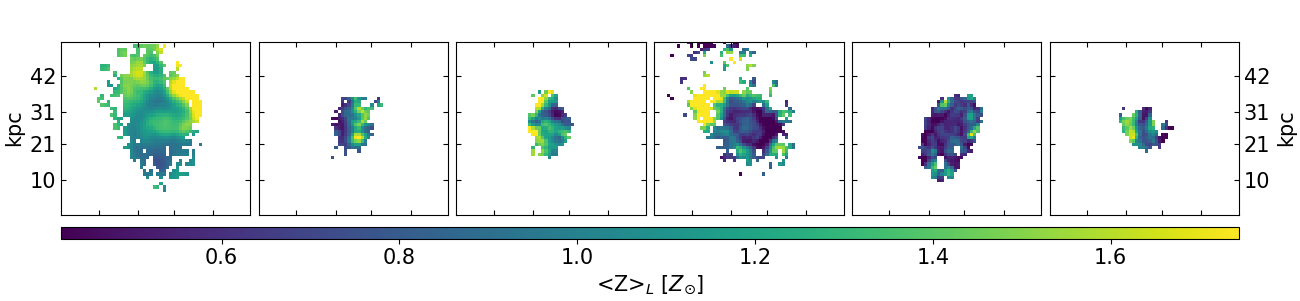}
    \caption{2D maps of luminosities, SFR surface densities, mean stellar ages and metallicities, for all six galaxies in our sample.}
    \label{maps}
\end{figure*}

As we have those measurements for each pixel of the galaxy, we can investigate the spatial distribution of the characteristics of the stellar populations and of the gas in the galaxies. We plot 2D morphological maps of some of the main quantities; that is, the integrated luminosity of the spectrum, the nebular extinction, mean stellar ages and metallicities weighted both by light and mass, and the SFRs computed through the population vector. Figure \ref{maps} shows the maps for the galaxy CAVITY52731.

\subsection{Integrated spectra}

Besides the methodology explained in previous sections for implementing the SPS in the spectra of each spaxel, we also computed the integrated spectra for each galaxy to measure global properties more easily and reliably. We summed all spectra for which the SPS were implemented, after the quality filters and {\sc urutau} implementation described in Sect. \ref{sps}. This way, we guaranteed a good S/N in the spectra ($\text{S/N} > 30$ in 4000-4060 \AA). Then we implemented {\sc fado} with them and could derive global measurements from the galaxies, such as total stellar mass and global mean stellar ages. Some measurements based on the synthesis of the integrated spectra are shown in Table \ref{tabela2}. Results shown in Table \ref{tabela3} and Figs. \ref{bpt_whan} and \ref{mass-assembly} were also obtained through the integrated spectra. This approach also allows for direct comparisons with the sample in \cite{costa-duarte16}, hereafter CD16.

\begin{table*}[]
\centering
\caption{Stellar population characteristics computed through {\sc fado} with the integrated spectra.}
\begin{tabular}{lllllllll}
\hline
Galaxy & $\log M_{\star} (\text{M}_{\odot})$ & $<\log t>_L$ & $<\log t>_M$ & $<Z>_L$ & $<Z>_M$ \\ \hline
VGS\_31a    & $9.47\pm0.06$  & $8.48\pm0.07$  & $9.20\pm0.18$  & $0.88\pm0.09$ & $1.47\pm0.25$ \\
VGS\_31b    & $9.58\pm0.03$  & $8.30\pm0.09$  & $9.18\pm0.07$  & $1.21\pm0.03$ & $1.81\pm0.14$ \\
VGS\_31c    & $8.96\pm0.46$  & $7.64\pm0.87$  & $8.35\pm1.36$  & $0.72\pm0.72$ & $0.53\pm0.72$ \\
CAVITY52731 & $10.94\pm0.01$ & $10.02\pm0.01$ & $10.11\pm0.01$ & $1.00\pm0.03$ & $1.07\pm0.05$ \\
CAVITY52732 & $9.76\pm0.11$  & $8.48\pm0.15$  & $9.51\pm0.24$  & $1.04\pm0.20$ & $1.66\pm0.37$ \\
CAVITY52733 & $9.85\pm0.04$  & $8.76\pm0.05$  & $9.73\pm0.08$  & $1.22\pm0.07$ & $2.29\pm0.03$ \\ \hline
\end{tabular}
\label{tabela2}
\end{table*}

CD16 analyses the stellar populations and ionization sources of 80 isolated galaxy triplets in the redshift range $0.04 \geq z \geq 0.1$, applying the SPS code {\sc starlight} in SDSS spectra. Although it uses single fiber spectra of the central parts, rather than IFU data, its approach is similar to ours when implementing SPS. And since it studies a large sample, it represents a good comparison base for this work. One caveat, however, is that the galaxies in that sample are at higher redshifts than the ones that are the focus of this work.

\section{Results and discussion}

\subsection{Emission lines}

\begin{figure*}
    \centering
    \includegraphics[width=\linewidth]{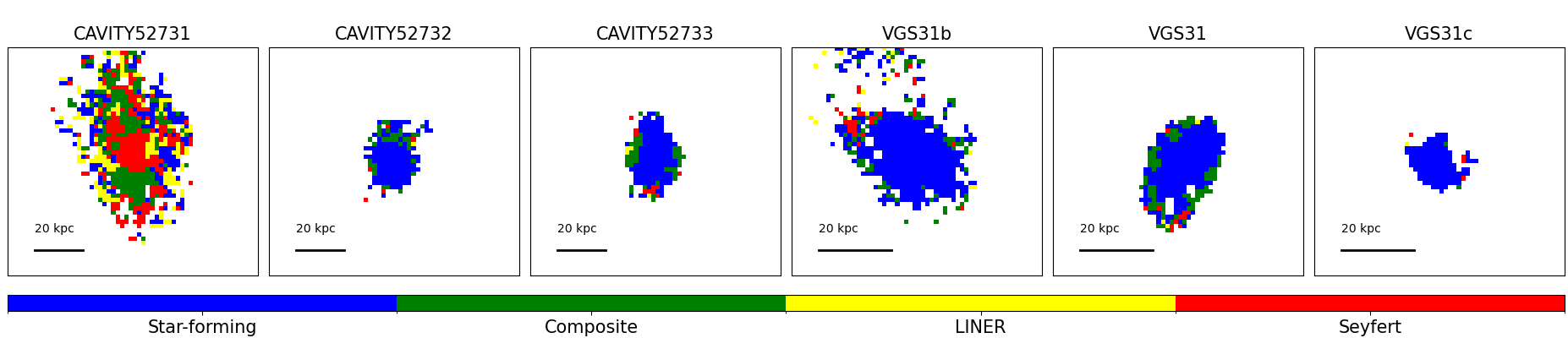}
    \caption{Maps of BPT classification for the galaxies. Blue is for star-forming spectra, green for composite, yellow for LINER-like emission, and red for Seyfert.}
    \label{bpt_maps}
\end{figure*}

\begin{figure*}
    \centering
    \includegraphics[width=\linewidth]{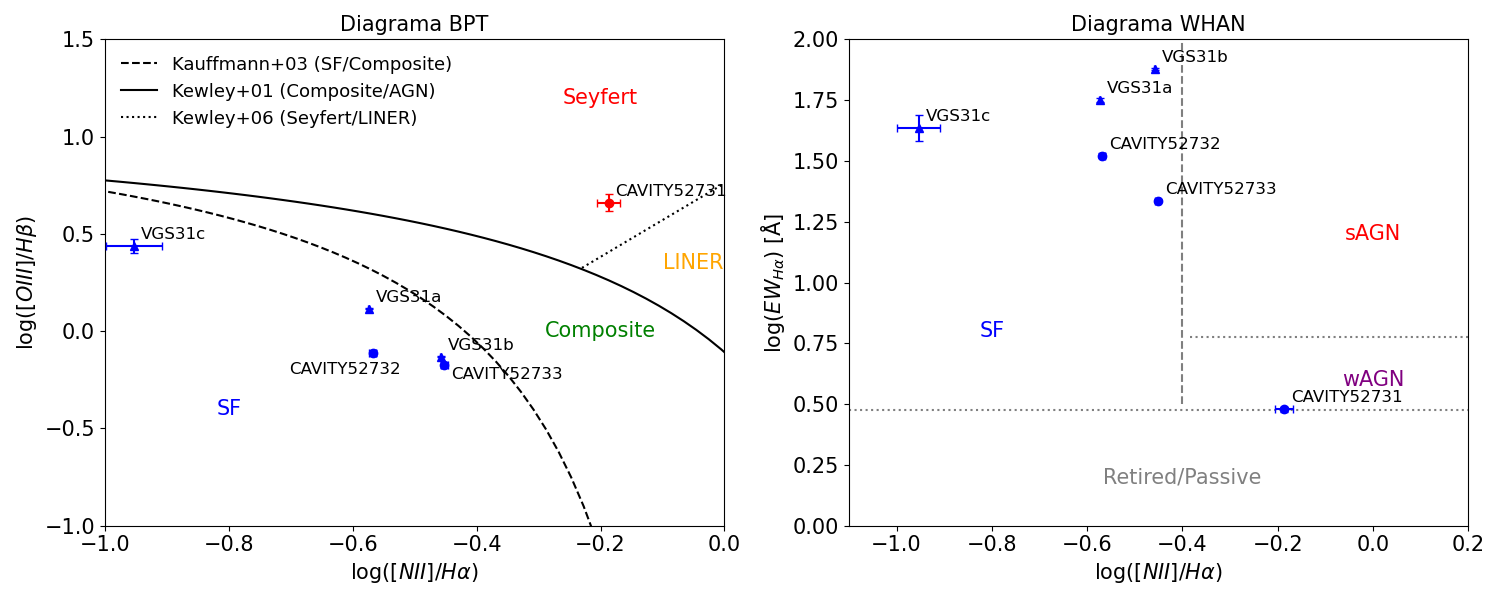}
    \caption{BPT diagram (left) and WHAN diagram (right) for the galaxies in this work, obtained from the integrated spectra.}
    \label{bpt_whan}
\end{figure*}

The gas of five of the galaxies that we investigate is ionized by recent star formation, as we verify in the maps of BPT classification shown in Fig. \ref{bpt_maps}. The only exception is CAVITY52731, the most massive in its system, which has its nebular emission powered mostly by Seyfert and composite sources. In fact, its main power source is classified as a Seyfert by the BPT diagram and as a weak active galactic nucleus (AGN) by the WHAN diagram, as is shown in Fig. \ref{bpt_whan} (unlike the maps of BPT classification for each spaxel, these diagrams were made using the integrated spectra). It is also the galaxy with the oldest stellar populations, being the only one in our sample with $<\log t>_L$ higher than 9. CAVITY52731 most likely is a massive ($M_{\star} \approx 10^{11} M_{\odot}$) quenched galaxy that hosts an AGN, while the rest are less massive ($M_{\star} < 10^{10} M_{\odot}$) active star-forming ones. Information about the emission lines and BPT and WHAN classifications of the sample is shown in Table \ref{tabela3}.

\begin{table*}[]
\centering
\caption{Emission lines fluxes and ratios, and BPT and WHAN classification computed through {\sc fado} with the integrated spectra (fluxes are in units of $\times10^{16}\text{erg/s/cm}^2$).}
\begin{tabular}{lllllllll}
\hline
Galaxy      & H$\alpha$ & H$\beta$ & log([\ion{N}{ii}]/H$\alpha$) & log([\ion{O}{iii}]/H$\beta$) & BPT & WHAN  \\ \hline
VGS\_31a    & $1569\pm2$ & $466\pm3$ & $-0.574\pm0.001$ & $0.116\pm0.003$   & SF      & SF \\
VGS\_31b    & $2849\pm3$ & $804\pm4$ & $-0.458\pm0.001$ & $-0.132\pm 0.003$ & SF      & SF \\
VGS\_31c    & $286\pm7$  & $74\pm6$  & $-0.953\pm0.045$ & $0.436\pm0.036$   & SF      & SF \\
CAVITY52731 & $188\pm4$  & $27\pm3$  & $-0.187\pm0.019$ & $0.661\pm0.044$   & Seyfert & wAGN     \\
CAVITY52732 & $313\pm2$  & $85\pm2$  & $-0.568\pm0.006$ & $-0.114\pm0.016$  & SF      & SF \\
CAVITY52733 & $202\pm1$  & $55\pm1$  & $-0.452\pm0.005$ & $-0.173\pm0.017$  & SF      & SF \\ \hline
\end{tabular}
\label{tabela3}
\end{table*}

According to CD16, most galaxies in isolated triplets are classified as AGN or passive in the WHAN diagram, even considering only the least massive systems, with only about 8\% of the galaxies being dominated by star formation. However, this result may also be biased by the fact that the spectra in CD16 are obtained with single fiber, covering the central parts of the galaxies. This way, galaxies that might be dominated by star formation in their outskirts may not be classified as mainly powered by star formation. Nevertheless, galaxies in our study are dominated by star formation, which might indicate they are among the rarest in terms of the ionization source. Also, VGS31c has a specially low value of $\log(\ion{N}{ii}/H_{\alpha})$.

\subsection{Mass assembly}

\begin{figure}
    \centering
    \includegraphics[width=\linewidth]{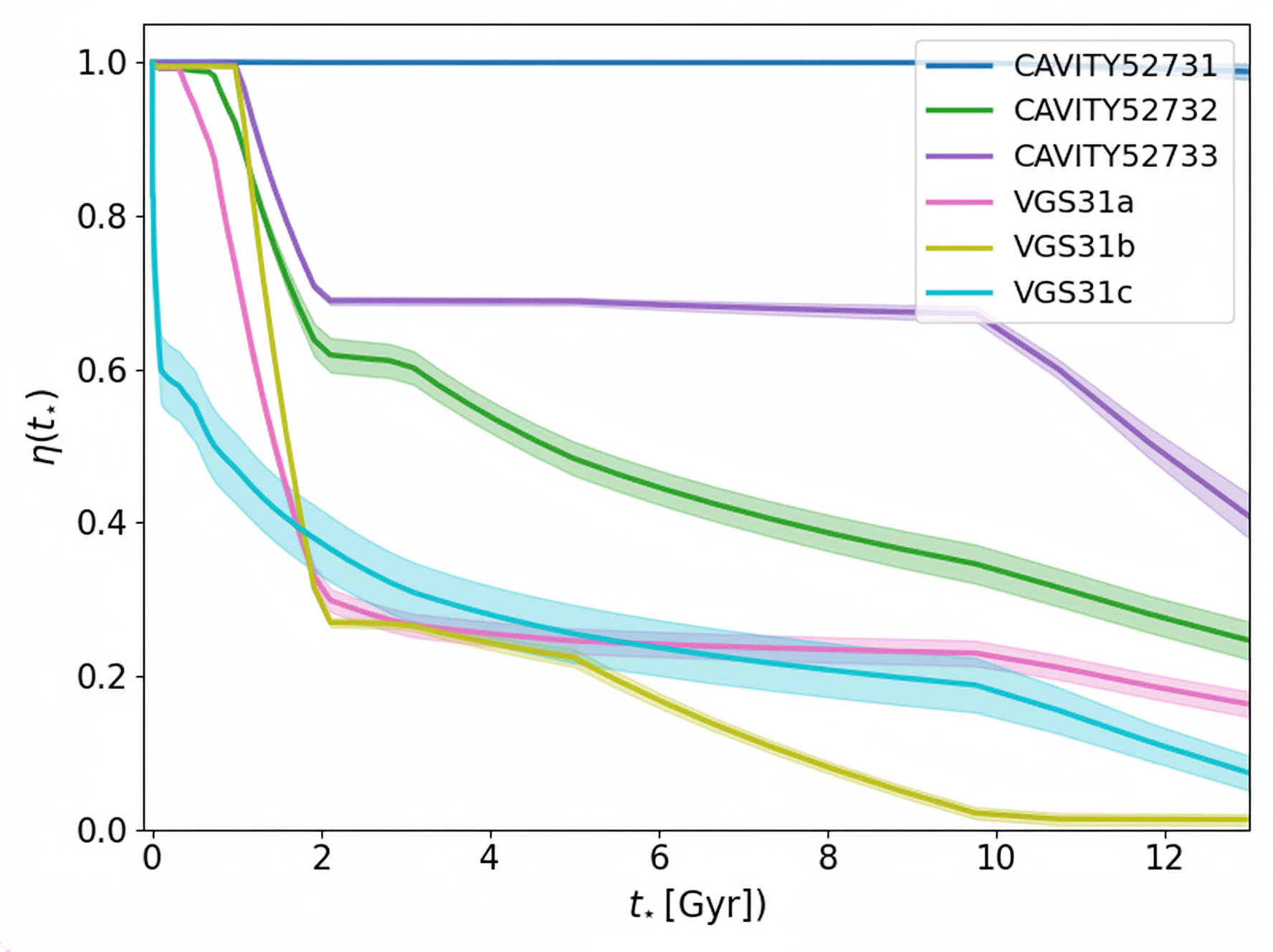}
    \caption{Stellar mass assembly function for the galaxies in triplets. The functions are the mean of the 100 different curves generated with Monte Carlo after perturbing the spectra and applying {\sc fado} in each of them. The shaded regions represent the error of the mean. These functions were obtained from the integrated spectra.}
    \label{mass-assembly}
\end{figure}

Except for CAVITY52731, all the other objects have stellar populations younger than 25 Myr that contribute significantly to the light (10\% - 40\%) but represent less than 1\% of the mass. We also computed the stellar mass assembly function \citep{asari07} for each galaxy, which is defined as
\begin{equation}
    \eta(t_{\star}) = \sum_i^{t<t_{\star}} \mu_{i} (t_i) \,.
\end{equation}
This is a cumulative function that grows from 0 to 1, starting at the oldest SSP in the spectral base and tracking what fraction of the stellar mass was due to stars formed up to a given look-back time. The computed curves are presented in Fig. \ref{mass-assembly}.

Here we also applied a simple Monte Carlo method, by perturbing the spectra of the galaxies 100 times, considering for each wavelength a Gaussian distribution with $\sigma$ equal to its error. Then we ran {\sc fado} in each new spectrum to reobtain the necessary quantities, such as total mass and population vectors. Then we had, for each galaxy, the mean of the 100 generated curves and the error of the mean.

\cite{dominguez-gomez23b} found that the SFHs at early times describe a bimodal distribution in the three large-scale environments (voids, filaments, walls and clusters), which allow us to classify the SFHs into two types: short-timescale SFH (ST-SFH) galaxies that formed a large fraction of their stellar mass (27\% on average) $\sim12.5$ Gyr ago, and long-timescale SFH (LT-SFH) galaxies that formed a lower fraction of their stellar mass ($<21.4\%$) than the ST-SFH galaxies 12.5 Gyr ago, but formed stars more uniformly over time. We can verify, based on Fig. \ref{mass-assembly}, that the three galaxies in VGS31 and CAVITY52732 have LT-SFH, while CAVITY52731 and CAVITY52733 have ST-SFH.

We verified that CAVITY52731 formed most of its mass at early epochs of the Universe ($t>10^{10} \mathrm{yr}$), while the others have significant star formation at more recent times. All the others present a rapid increase in their mass in the last 2 Gyr. Before that, their star formation was more continuous. This is in agreement with the previous results (see Figs. \ref{bpt_maps} and \ref{bpt_whan}) and evidence that only CAVITY52731 is a passive galaxy in those systems. Although CAVITY52733 is classified as ST-SFH by the criteria of the mass formed 12.5 Gyr ago, the increase in the SFR at earlier times is more similar to the behavior of an LT-SFH. It is not totally in accordance with the general behavior found for cluster galaxies (and other environments as well), which assemble on average 30\% of their mass at early times and decrease their SFR later in their lives. This may be indicative of a more complex evolutionary path than non-triplet galaxies. In fact, \cite{dominguez-gomez23b} presents the mass assembly function for the galaxies in CAVITY, grouping by LT-SFH and ST-SFH, revealing their general behavior, and their functions are smooth through time, as well as the functions for two examples of CAVITY galaxies shown in the same work. Those facts might indicate that the great increase in SFR at late times, as we see in our sample, is not common for void galaxies.

The mass assembly function for CAVITY52731 shows no star formation in the last 10 Gyr, evidencing a fast mass assembly common for massive and quiescent galaxies. However, we must consider that the spectra of old SSPs with the same metallicity are nearly indistinguishable. For ages on the gigayear scale, the spectra show little evolution with time. In our base, we chose only three ages to represent the old populations: 5, 10, and 13 Gyr (for ages lower than 1 Gyr, the age resolution of the base is much higher). That means that the galaxy could, in fact, have star formation at times between those large intervals, but the SPS codes do not differentiate them much. In fact, verifying the population vectors for the various results in the Monte Carlo for CAVITY52731, there is in fact populations with $t_{\star} = 5\, \text{Gyr}$ or $t_{\star} = 1\, \text{Gyr}$, although they never represent more than 0.2\% of the mass. We also detect populations with $t_{\star} = 5\, \text{Myr}$ that are responsible for less than 5\% of the light.

We also leveraged the Monte Carlo implementation in the SPS to compute the stellar masses of the galaxies, which are shown in Table \ref{tabela2}. Both triplets are among the least massive triplets, being below the 33.3\% percentile of the stellar mass distribution from CD16 ($\log M_{\star} < 11.52$). This may already be expected due to the fact that galaxies in voids are, in general, less massive than galaxies in filaments, walls, and clusters. VGS31 has a total stellar mass of $\log M_{\star} = 9.86$ and CAVITY5273X has $\log M_{\star} = 10.99$. In that sense, VGS31 is a particularly unique triplet, since it is formed by three dwarf galaxies. The least massive systems in CD16 have $\log M_{\star} (\text{M}_{\odot})\sim 11.0$. The uncertainties in the masses are quite small (up to $\sim 0.1 \text{dex}$ in the logarithmic space), except for VGS31c, which has an uncertainty of $\sim 0.5$.

\subsection{Mass-metallicity relation}

Finally, we compared the stellar metallicities of the six galaxies to the mass-metallicity relation (MMR) extracted from \cite{dominguez-gomez23a} for void galaxies. In that work, the stellar metallicities were obtained through a nonparametric full spectral fitting analysis applied to SDSS-DR7 spectra using the \textsc{pPXF} \citep{cappellari17} and \textsc{steckmap} \citep{ocvirk06a,ocvirk06b} codes. They modeled the spectra in the 3750–5450 Å range, combining single stellar population (SSP) templates from the E-MILES library and assuming a Kroupa IMF. The stellar line-of-sight velocity distribution and gas emission lines were first fit with \textsc{pPXF}; the emission was subtracted, and the stellar population parameters were then derived with \textsc{steckmap} assuming the fixed kinematics from \textsc{pPXF}. The SSP metallicities cover $2.27 \leq [\text{M/H}] \leq 0.40$ dex, where $[\text{M/H}] = \log (Z/Z_{\odot})$.

\cite{dominguez-gomez23a} computes the stellar metallicity in the galaxies within their central 3 arcsec. The sample used to represent the void galaxies was the CAVITY sample as well, and since each fiber of PPAK has a diameter of 2.7 arcsec, we used the spectrum of the central spaxel in the galaxy to derive the metallicity. Following \cite{dominguez-gomez23a}, we computed the metallicities as
\begin{equation}
    [\text{M/H}]_M = \sum_{j=1}^{N_{\star}} \mu_j \log (Z_j/Z_{\odot}) \,,
\end{equation}
where $Z_j$ is the metallicity of the j-th SSP in the base and $\mu_j$ is the mass fraction of that population.

\begin{figure*}
    \centering
    \includegraphics[width=\linewidth]{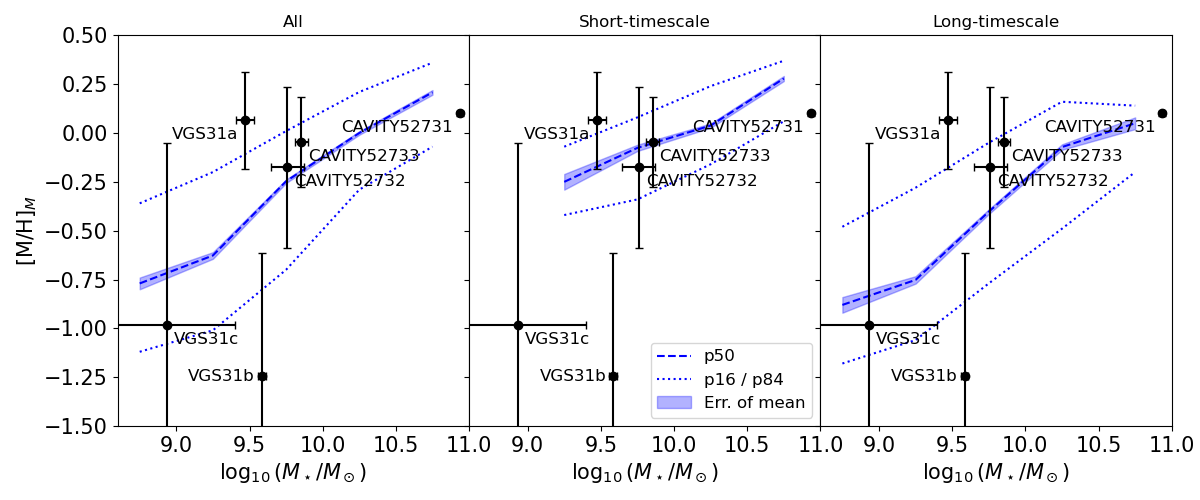}
    \caption{Mass-metallicity relation for the galaxies in our sample. In blue we have the MMR for void galaxies, taken from \cite{dominguez-gomez23a}. The dashed lines are the mean values, the shaded regions are the error of the mean, and the dotted lines mark the $1\sigma$ distribution in metallicity. The MMR is shown for three different samples: all void galaxies (from CAVITY), ST-SFH void galaxies, and LT-SFH void galaxies. The black dots indicate the masses and central metallicities of the galaxies in our sample. The points are the mean of the 100 different points generated for each galaxy with Monte Carlo after perturbing the spectra and applying {\sc fado} in each of them. The error bars indicate the standard deviation of the Monte Carlo generated sample.}
    \label{mass-metalicity}
\end{figure*}

The MMR is shown in Fig. \ref{mass-metalicity}. The galaxies in CAVITY5273X follow the MMR quite well, being inside one standard deviation ($\sigma$) of the distribution. On the other hand, VGS31a and VGS31b are outliers, as they have $[\text{M/H}]_M$ outside $1\sigma$ of the distribution. This could be due to their more complex evolutionary path, since they are embedded in an \ion{H}{i} filament and are interacting. Besides the fact that all of them are dwarf galaxies, VGS31a shows signatures of a past merger, and VGS31a shows indications of tidal interactions with the other galaxy or accretion from filamentary material \citep{beygu13}. In fact, all three galaxies could accrete filamentary material, which, together with the interactions, might trigger or keep star formation. This process enriches the gas and stars subsequently formed, along with the fact that the gas in filaments is generally less metal-rich compared to the galaxies.

In the case of CAVITY5273X, the galaxies also seem to be interacting with one another, although we cannot be sure since we do not have more details on their dynamics. Also, it is not embedded in a filament, so it appears to be a less complex scenario, but the interactions could cause different episodes of star formation in different regions of the galaxies, which may lead to complex stellar metallicities. We also verify that metallicities can vary significantly across different locations within those galaxies, as is shown in the maps in Fig. \ref{maps}, rather than analyzing only the integrated spectra.

\cite{dominguez-gomez23a} derived MMRs separately for those two different SFHs (LT-SFH and ST-SFH) as well, which are also shown in Fig. \ref{mass-metalicity}. CAVITY52732 and CAVITY52731 fall inside the $1\sigma$ distribution for both relations, while the galaxies VGS31b and VGS31c are still outliers in both distributions.

\subsection{Morphology}
\label{morphology}

All galaxies in both triplets present asymmetric structures that indicate interactions. In order to complement the stellar population and ionized gas analysis performed in the previous sections, we computed a series of morphological parameters, which allowed us to better understand the structure of those objects.

Here we made use of the code {\sc morfometryka} \citep{morfometryka}. It consists of a standalone application to automatically perform all the structural and morphometric measurements over a galaxy image. Between them are the original and modified parameters of the concentration, asymmetry, smoothness, Gini, and M20 (CASGM) system, presented in \cite{abraham94,abraham96,conselice00,lotz04}, in addition to Sérsic indexes \citep{sersic68}, and the new parameters entropy and spirality. The most recent version of {\sc morfometryka} also provides the curvature of the brightness profile with {\sc kurvature} \citep{lucatelli19}, which is a powerful tool for probing the presence of multiple components in galaxies.

\begin{figure*}
    \centering
    \includegraphics[width=0.95\linewidth]{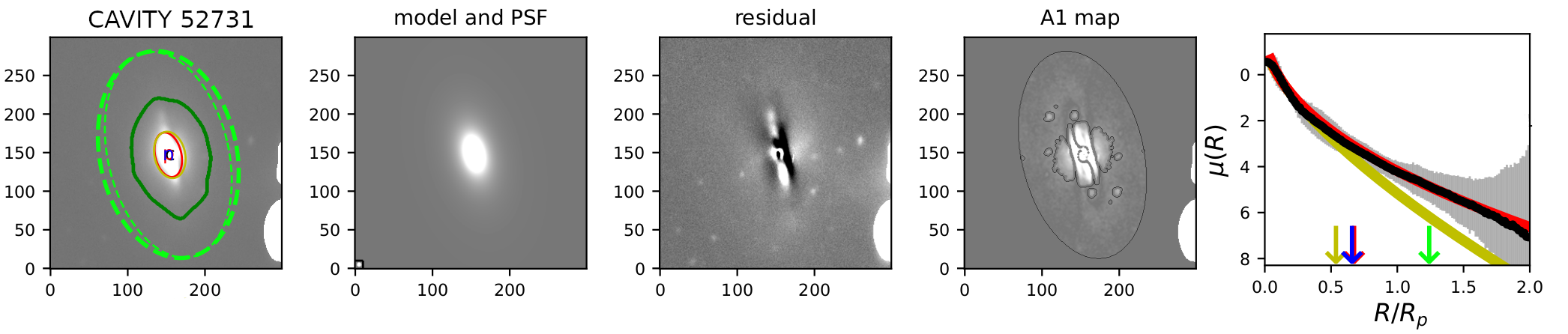}
    \includegraphics[width=0.95\linewidth]{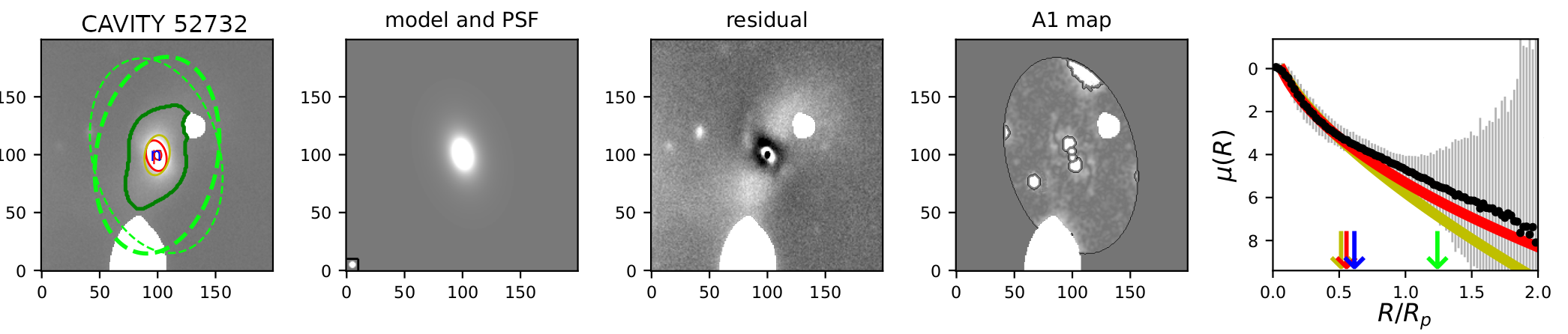}
    \includegraphics[width=0.95\linewidth]{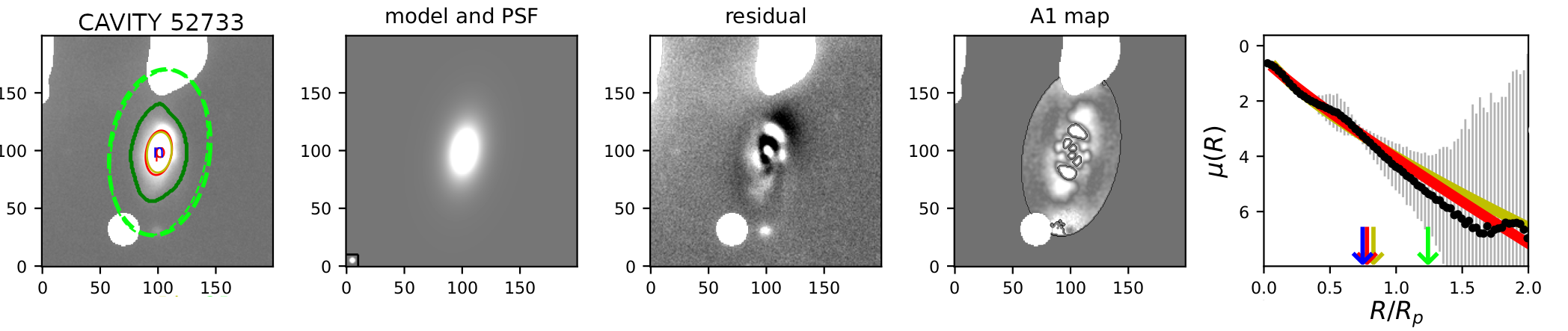}
    \includegraphics[width=0.95\linewidth]{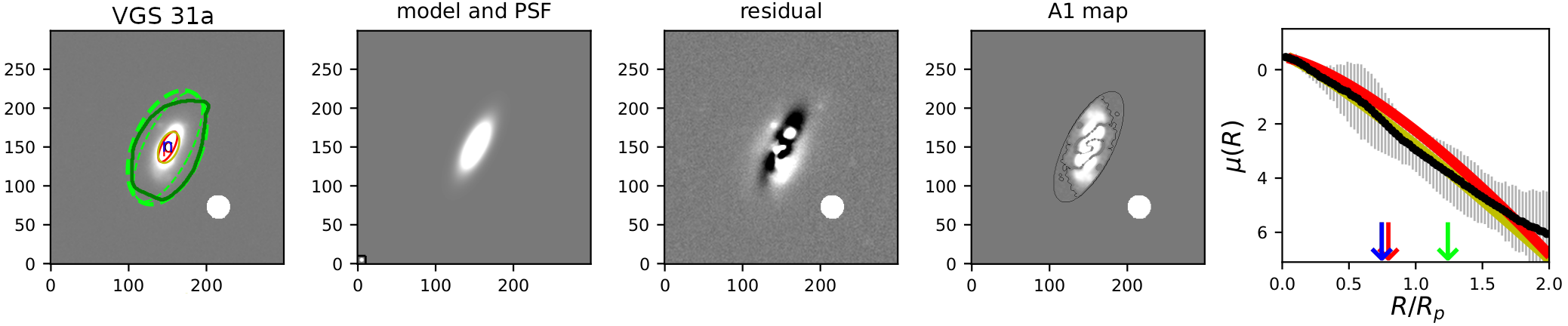}
    \includegraphics[width=0.95\linewidth]{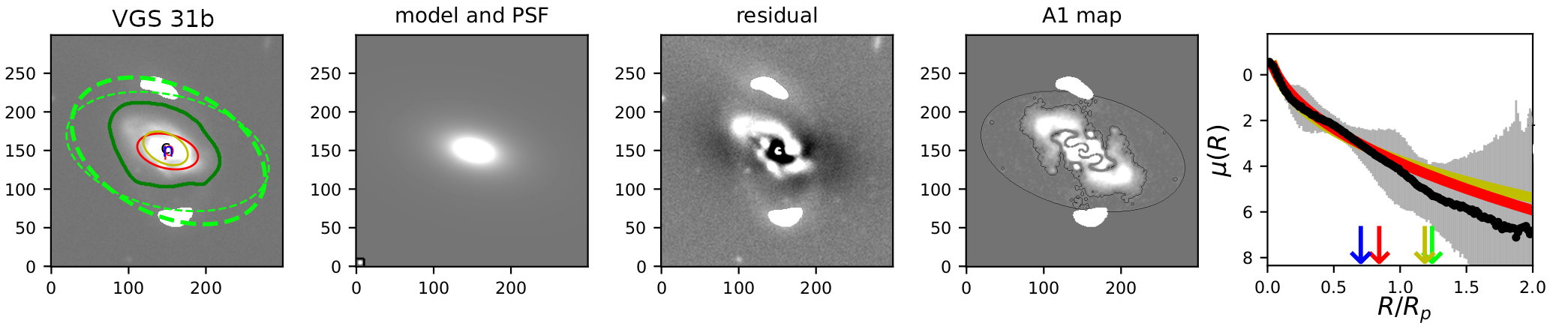}
    \includegraphics[width=0.95\linewidth]{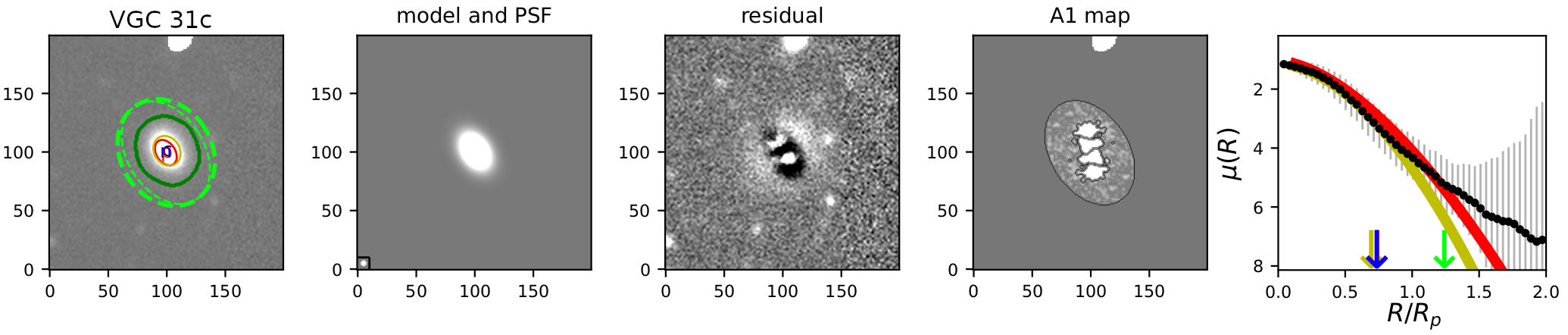}
    \caption{\textsc{Morfometryka} output for CAVITY 52731 (upper row), CAVITY 52732 (middle), and CAVITY 52733 (lower). In the columns, from left to right: original image in $g$ band ($\rm arcsinh$ scale); single Sérsic model fit and PSF (lower left insert); single Sérsic residual; asymmetry map; and brightness profile (black dots) and 1D (yellow) and 2D (red) derived Sérsic models. In the leftmost panels, the dark green line is the initial segmentation region; the neon green lines are the two Petrosian radius regions (obtained both by fitting Sérsic and by image momenta); the red line corresponds to the Sérsic $R_n$. In the rightmost panels, black dots and bars are measurements and associated uncertainties; the yellow line is the Sérsic fit to the 1D profile; the red line is the profile but with the parameters obtained from the image (2D) fits. }
    \label{mfmtk-cavity}
\end{figure*}

Figure \ref{mfmtk-cavity} shows the morphology decompositions of the galaxies in CAVITY5273X and VGS31, alongside with the models, residuals and A1 map computed by {\sc morfometryka}, as well as their luminosity profiles, and 1D and 2D Sérsic profiles. The images used are from the $g$ band of the DESI Legacy Survey \citep{dey19}. A1 is measured as defined by \cite{abraham96}, by subtracting the rotated galaxy image ($I_{\pi}$) from the original galaxy image ($I$) within the Petrosian radius ($R_p$) and without subtracting the sky, following

\begin{equation}
    A_1 = \frac{\text{abs}(I-I_{\pi})}{2I}\,.
\end{equation}

The residuals and A1 maps evidence the asymmetry present in all galaxies in the two triplets. CAVITY52732 and VGS31b are probably the most eye-catching in the sample due to their clear disturbance. The first presents two highlighted tidal tails, one extended to the northwest and another to the southeast, and the last presents a ring-like structure to the south and a long tail to the northeast. Such features are relatively easy to identify just by looking at the RGB images of the systems. But unfortunately, such structures are mostly too faint to have sufficient S/N and be synthesized by our SPS method.

CAVITY52733 also has an extended emission to the south, and VGS31a has one to the southeast. VGS31c, being tinier, may be the most difficult one in which to identify any disturbance. The only one that seems undisturbed or mildly perturbed is CAVITY52731, maybe because the mass of its companions is too low in proportion to significantly affect the orbits of the stars, and they have not merged yet.

Regarding the surface brightness profile, we see a large-scale structure that has a convex shape in the surface brightness profile in most of the galaxies. However, {\sc morfometryka} cannot fit most of the details that stem from the irregular structure, so a more accurate profile may deviate even more than a single Sérsic component profile. The insufficiency of single-component Sérsic models to describe the surface brightness is one more piece of evidence of the disturbed morphology of those galaxies.

The only galaxy that is well represented by a 2D Sérsic model is CAVITY52731, which is a passive galaxy and dominates its system in mass. Its profile is close to an elliptical galaxy with $n_{2D} \sim 2$, although it presents an obscured stripe parallel to its semimajor axis (SMA), which resembles the extinction by the dust in a disk seen edge-on. This could indicate that this galaxy has an S0 morphology. CAVITY52732, which has two tidal tails, is relatively well modeled by a Sérsic model also with $n_{2D} \sim 2$ up to radii $0.5R_p$, but presents a less concentrated brightness for larger distances. CAVITY52733 has a structure closer to a disk ($n_{2D} \sim 1$) up to radii of $R \sim 0.75R_p$.

When it comes to the triplet in the filament, they seem to deviate even more than their 2D Sérsic models. The profiles for both VGS31a and VGS31c transition from a concave to a convex curve around $R \sim 0.8R_p$. VGS31b, the one with an arc to the south and a long tail to the north, has a concentrated light in the center, after transitioning to a more concave curve up to $R \sim 1R_p$, and after returning to a convex profile. The tail and the ring-like structures are likely caused by a past minor merger incident with a low-mass galaxy \citep{mihos&hernquist94,mihos&hernquist96,duc&renaud13}.

A multiwavelength study of VGS31 triplet was performed by \cite{beygu13}, combining observations of the optical, $H_{\alpha}$, NUV and FUV, and CO(1-0). That work finds that the tail and arc structures in VGS31b are devoid of star formation, in opposition to its central regions, which characterizes a starburst. The galaxy presents a bar, and its kinematics are indicative of fast-rotating inner structure and streaming motions. VGS31a also has disturbed internal kinematics and an enhanced SFR, which could be explained by a tidal interaction with VGS31b or by the accretion of the filamentary material. VGS31c also has an enhanced SFR and disturbed kinematics, but it is difficult to further investigate the possible scenarios because of the low S/N.

\section{Summary and conclusions}

In this work, we have presented a spatially resolved analysis of six galaxies in two triplets residing within cosmological voids, using integral field spectroscopy from the CAVITY survey. Our results reinforce the view that, despite the overall low-density conditions in voids, galaxies therein can exhibit rich and diverse evolutionary histories driven by internal mechanisms and local interactions. The main conclusions of our study are:

\begin{itemize}
    \item CAVITY52731 is markedly different from the rest of the sample, showing signs of early mass assembly, while the remaining galaxies display prominent young stellar populations and active star formation.
    
    \item Morphological features such as tidal tails, arcs, and asymmetric light distributions are common in both triplets, indicating that galaxy interactions and minor mergers are actively shaping these systems even in void regions.
    
    \item Emission-line diagnostics confirm that most galaxies are powered by recent star formation, with CAVITY52731 being the only exception, exhibiting AGN-like signatures in both BPT and WHAN diagrams.

    \item Stellar mass assembly histories reveal that while some galaxies formed most of their mass early in the Universe’s history, others underwent significant recent star formation, reflecting diverse and nonuniversal evolutionary paths. While one of the star-forming galaxies (CAVITY52733) has an ST-SHF and the other four have an LT-SFH, all of them present accelerated mass assembly in the last 2 Gyr.

    \item The stellar MMR reveals contrasting behavior between the two systems. While CAVITY5273X galaxies follow the MMR expected for void galaxies, the VGS31 galaxies appear as significant outliers, particularly in terms of metallicity. This may indicate the influence of filamentary accretion.

    \item Sérsic profile fits and residuals demonstrate that most galaxies deviate significantly from regular morphologies, further confirming their disturbed nature and supporting scenarios involving interactions and accretion.

\end{itemize}

These findings emphasize that galaxy evolution in voids is not a passive process. Instead, local interactions and small-scale structures, such as triplets and filaments, play a critical role in driving stellar and morphological transformations. Furthermore, deviations from standard relations such as the MMR provide valuable insights into how environmental context can shape the chemical evolution of galaxies.

\begin{acknowledgements}

    Based on observations collected at the Centro Astron\'omico Hispano en Andaluc\'ia (CAHA) at Calar Alto, operated jointly by Junta de Andaluc\'ia and Consejo Superior de Investigaciones Cient\'ificas (IAACSIC). The CAVITY project acknowledges financial support from projects PID2020-113689GB-I00 and PID2023-149578NB-I00, financed by MCIN/AEI/10.13039/501100011033 and FEDER/UE; from FEDER/Junta de Andaluc\'ia-Consejería de Transformaci\'on Econ\'omica, Industria, Conocimiento y Universidades/Proyecto A-FQM-510-UGR20; from grant AST22\_4.4 financed by Junta de Andalucía-Consejería de Universidad, Investigaci\'on e Innovación and Gobierno de España and European Union-NextGenerationEU; and from projects P20\_00334 and FQM 108 financed by the Junta de Andalucia. GMA acknowledges financial support Coordenação de Aperfeiçoamento de Pessoal de Nível Superior (CAPES Proj. 0001) and the Programa de Pós-Graduação em Física (PPGFis) at UFRGS. MAF acknowledges support from the Emergia program (EMERGIA$20\_38888$) from Consejería de Universidad, Investigación e Innovación de la Junta de Andalucía. BB acknowledges financial support from the Grant AST22-4.4, funded by Consejería de Universidad, Investigación e Innovación and Gobierno de España and Unión Europea – NextGenerationEU, and by the research projects PID2020-113689GB-I00 and PID2023-149578NB-I00 financed by MCIN/AEI/10.13039/501100011033. BB acknowledges financial support from the Grant AST22-4.4, funded by Consejería de Universidad, Investigación e Innovación and Gobierno de España and Unión Europea – NextGenerationEU, and by the research projects PID2020-113689GB-I00 and PID2023-149578NB-I00 financed by MCIN/AEI/10.13039/501100011033. 
  ALCS acknowledges support from FAPERGS (grants 23/2551-0001832-2 and 24/2551-0001548-5), CNPq (grants 314301/2021-6, 312940/2025-4, 445231/2024-6, and 404233/2024-4), and CAPES (grant 88887.004427/2024-00).
    RR acknowledges support from   CNPq(445231/2024-6,311223/2020-6, 404238/2021-1, and 310413/2025-7), FAPERGS (19/1750-2 and 24/2551-0001282-6) and CAPES(88881.109987/2025-01).
      
\end{acknowledgements}

\bibliographystyle{aa}
\bibliography{refs}

@ARTICLE{Zhang2022,
       author = {{Zhang}, Yikun and {de Souza}, Rafael S. and {Chen}, Yen-Chi},
        title = "{SCONCE: a cosmic web finder for spherical and conic geometries}",
      journal = {\mnras},
         year = 2022,
        month = nov,
       volume = {517},
       number = {1},
        pages = {1197-1217},
          doi = {10.1093/mnras/stac2504},
archivePrefix = {arXiv},
       eprint = {2207.07001},
 primaryClass = {astro-ph.CO},
       adsurl = {https://ui.adsabs.harvard.edu/abs/2022MNRAS.517.1197Z}
}

@ARTICLE{Moews2021,
       author = {{Moews}, Ben and {Schmitz}, Morgan A. and {Lawler}, Andrew J. and {Zuntz}, Joe and {Malz}, Alex I. and {de Souza}, Rafael S. and {Vilalta}, Ricardo and {Krone-Martins}, Alberto and {Ishida}, Emille E.~O. and {COIN Collaboration}},
        title = "{Ridges in the Dark Energy Survey for cosmic trough identification}",
      journal = {\mnras},
         year = 2021,
        month = jan,
       volume = {500},
       number = {1},
        pages = {859-870},
          doi = {10.1093/mnras/staa3204},
archivePrefix = {arXiv},
       eprint = {2005.08583},
 primaryClass = {astro-ph.CO},
       adsurl = {https://ui.adsabs.harvard.edu/abs/2021MNRAS.500..859M}
}

@ARTICLE{fado,
       author = {{Gomes}, J.~M. and {Papaderos}, P.},
        title = "{Fitting Analysis using Differential evolution Optimization (FADO):. Spectral population synthesis through genetic optimization under self-consistency boundary conditions}",
      journal = {\aap},
         year = 2017,
        month = jul,
       volume = {603},
          eid = {A63},
        pages = {A63},
          doi = {10.1051/0004-6361/201628986},
archivePrefix = {arXiv},
       eprint = {1704.03922},
 primaryClass = {astro-ph.GA},
       adsurl = {https://ui.adsabs.harvard.edu/abs/2017A&A...603A..63G}
}

@ARTICLE{fado2,
       author = {{Cardoso}, Leandro S.~M. and {Gomes}, Jean Michel and
         {Papaderos}, Polychronis},
        title = "{Self-consistent population spectral synthesis with FADO. I. The importance of nebular emission in modelling star-forming galaxies}",
      journal = {\aap},
         year = 2019,
        month = feb,
       volume = {622},
          eid = {A56},
        pages = {A56},
          doi = {10.1051/0004-6361/201833438},
archivePrefix = {arXiv},
       eprint = {1901.10776},
 primaryClass = {astro-ph.GA},
       adsurl = {https://ui.adsabs.harvard.edu/abs/2019A&A...622A..56C}
}

@ARTICLE{BC03,
       author = {{Bruzual}, G. and {Charlot}, S.},
        title = "{Stellar population synthesis at the resolution of 2003}",
      journal = {\mnras},
         year = 2003,
        month = oct,
       volume = {344},
       number = {4},
        pages = {1000-1028},
          doi = {10.1046/j.1365-8711.2003.06897.x},
archivePrefix = {arXiv},
       eprint = {astro-ph/0309134},
 primaryClass = {astro-ph},
       adsurl = {https://ui.adsabs.harvard.edu/abs/2003MNRAS.344.1000B}
}

@ARTICLE{chabrier,
       author = {{Chabrier}, Gilles},
        title = "{Galactic Stellar and Substellar Initial Mass Function}",
      journal = {\pasp},
         year = "2003",
        month = "Jul",
       volume = {115},
       number = {809},
        pages = {763-795},
          doi = {10.1086/376392},
archivePrefix = {arXiv},
       eprint = {astro-ph/0304382},
 primaryClass = {astro-ph},
       adsurl = {https://ui.adsabs.harvard.edu/abs/2003PASP..115..763C}
}

@ARTICLE{dametto14,
       author = {{Dametto}, N.~Z. and {Riffel}, R. and {Pastoriza}, M.~G. and {Rodr{\'\i}guez-Ardila}, A. and {Hernandez-Jimenez}, J.~A. and {Carvalho}, E.~A.},
        title = "{Probing the circumnuclear stellar populations of starburst galaxies in the near-infrared}",
      journal = {\mnras},
         year = 2014,
        month = sep,
       volume = {443},
       number = {2},
        pages = {1754-1778},
          doi = {10.1093/mnras/stu1243},
archivePrefix = {arXiv},
       eprint = {1406.6345},
 primaryClass = {astro-ph.GA},
       adsurl = {https://ui.adsabs.harvard.edu/abs/2014MNRAS.443.1754D}
}

@ARTICLE{azevedo23,
       author = {{Azevedo}, Gabriel M. and {Chies-Santos}, Ana L. and {Riffel}, Rog{\'e}rio and {Gomes}, Jean M. and {Lassen}, Augusto E. and {Benedetti}, Jo{\~a}o P.~V. and {de Souza}, Rafael S. and {Xu}, Quanfeng},
        title = "{Spatially resolved self-consistent spectral modelling of jellyfish galaxies from MUSE with FADO: trends with mass and stripping intensity}",
      journal = {\mnras},
         year = 2023,
        month = aug,
       volume = {523},
       number = {3},
        pages = {4680-4692},
          doi = {10.1093/mnras/stad1641},
archivePrefix = {arXiv},
       eprint = {2306.00049},
 primaryClass = {astro-ph.GA},
       adsurl = {https://ui.adsabs.harvard.edu/abs/2023MNRAS.523.4680A}
}

@article{starlight,
    author = {Cid Fernandes, Roberto and Mateus, Abílio and Sodré, Laerte and Stasińska, Grażyna and Gomes, Jean M.},
    title = "{Semi-empirical analysis of Sloan Digital Sky Survey galaxies – I. Spectral synthesis method}",
    journal = {Monthly Notices of the Royal Astronomical Society},
    volume = {358},
    number = {2},
    pages = {363-378},
    year = {2005},
    month = {04},
    issn = {0035-8711},
    doi = {10.1111/j.1365-2966.2005.08752.x},
    url = {https://doi.org/10.1111/j.1365-2966.2005.08752.x},
    eprint = {http://oup.prod.sis.lan/mnras/article-pdf/358/2/363/18422924/358-2-363.pdf},
}

@ARTICLE{cavity,
       author = {{P{\'e}rez}, I. and {Verley}, S. and {S{\'a}nchez-Menguiano}, L. and {Ruiz-Lara}, T. and {Garc{\'\i}a-Benito}, R. and {Duarte Puertas}, S. and {Jim{\'e}nez}, A. and {Dom{\'\i}nguez-G{\'o}mez}, J. and {Espada}, D. and {Peletier}, R.~F. and {Rom{\'a}n}, J. and {Rodr{\'\i}guez}, M.~I. and {S{\'a}nchez Alarc{\'o}n}, P. and {Argudo-Fern{\'a}ndez}, M. and {Torres-R{\'\i}os}, G. and {Bidaran}, B. and {Alc{\'a}zar-Laynez}, M. and {van de Weygaert}, R. and {S{\'a}nchez}, S.~F. and {Lisenfeld}, U. and {Zurita}, A. and {Florido}, E. and {van der Hulst}, J.~M. and {Bl{\'a}zquez-Calero}, G. and {Villalba-Gonz{\'a}lez}, P. and {del Moral-Castro}, I. and {Lugo-Aranda}, A. and {Walo-Mart{\'\i}n}, D. and {Conrado}, A. and {Gonz{\'a}lez Delgado}, R. and {Falc{\'o}n-Barroso}, J. and {Ferr{\'e}-Mateu}, A. and {Hern{\'a}ndez-S{\'a}nchez}, M. and {Awad}, P. and {Kreckel}, K. and {Courtois}, H. and {Espada-Miura}, R. and {Rela{\~n}o}, M. and {Galbany}, L. and {S{\'a}nchez-Bl{\'a}zquez}, P. and {P{\'e}rez-Montero}, E. and {S{\'a}nchez-Portal}, M. and {Bongiovanni}, A. and {Planelles}, S. and {Quilis}, V. and {Aubert}, M. and {Guinet}, D. and {Pomar{\'e}de}, D. and {Weijmans}, A.~M. and {Raj}, M.~A. and {Arag{\'o}n-Calvo}, M.~A. and {Azzaro}, M. and {Bergond}, G. and {Blazek}, M. and {Cikota}, S. and {Fern{\'a}ndez-Mart{\'\i}n}, A. and {Gardini}, A. and {Guijarro}, A. and {Hermelo}, I. and {Mart{\'\i}n}, P. and {Vico Linares}, J.~I.},
        title = "{CAVITY, Calar Alto Void Integral-field Treasury surveY and project extension}",
      journal = {\aap},
         year = 2024,
        month = may,
       volume = {689},
          eid = {A213},
        pages = {A213},
          doi = {10.1051/0004-6361/202449749},
archivePrefix = {arXiv},
       eprint = {2405.04217},
 primaryClass = {astro-ph.GA},
       adsurl = {https://ui.adsabs.harvard.edu/abs/2024arXiv240504217P}
}

@ARTICLE{pmas,
       author = {{Roth}, Martin M. and {Kelz}, Andreas and {Fechner}, Thomas and {Hahn}, Thomas and {Bauer}, Svend-Marian and {Becker}, Thomas and {B{\"o}hm}, Petra and {Christensen}, Lise and {Dionies}, Frank and {Paschke}, Jens and {Popow}, Emil and {Wolter}, Dieter and {Schmoll}, J{\"u}rgen and {Laux}, Uwe and {Altmann}, Werner},
        title = "{PMAS: The Potsdam Multi-Aperture Spectrophotometer. I. Design, Manufacture, and Performance}",
      journal = {\pasp},
         year = 2005,
        month = jun,
       volume = {117},
       number = {832},
        pages = {620-642},
          doi = {10.1086/429877},
archivePrefix = {arXiv},
       eprint = {astro-ph/0502581},
 primaryClass = {astro-ph},
       adsurl = {https://ui.adsabs.harvard.edu/abs/2005PASP..117..620R}
}

@ARTICLE{ppak,
       author = {{Kelz}, Andreas and {Verheijen}, Marc A.~W. and {Roth}, Martin M. and {Bauer}, Svend M. and {Becker}, Thomas and {Paschke}, Jens and {Popow}, Emil and {S{\'a}nchez}, Sebastian F. and {Laux}, Uwe},
        title = "{PMAS: The Potsdam Multi-Aperture Spectrophotometer. II. The Wide Integral Field Unit PPak}",
      journal = {\pasp},
         year = 2006,
        month = jan,
       volume = {118},
       number = {839},
        pages = {129-145},
          doi = {10.1086/497455},
archivePrefix = {arXiv},
       eprint = {astro-ph/0512557},
 primaryClass = {astro-ph},
       adsurl = {https://ui.adsabs.harvard.edu/abs/2006PASP..118..129K}
}

@ARTICLE{astropy,
       author = {{Astropy Collaboration} and {Price-Whelan}, Adrian M. and {Lim}, Pey Lian and {Earl}, Nicholas and {Starkman}, Nathaniel and {Bradley}, Larry and {Shupe}, David L. and {Patil}, Aarya A. and {Corrales}, Lia and {Brasseur}, C.~E. and {N{\"o}the}, Maximilian and {Donath}, Axel and {Tollerud}, Erik and {Morris}, Brett M. and {Ginsburg}, Adam and {Vaher}, Eero and {Weaver}, Benjamin A. and {Tocknell}, James and {Jamieson}, William and {van Kerkwijk}, Marten H. and {Robitaille}, Thomas P. and {Merry}, Bruce and {Bachetti}, Matteo and {G{\"u}nther}, H. Moritz and {Aldcroft}, Thomas L. and {Alvarado-Montes}, Jaime A. and {Archibald}, Anne M. and {B{\'o}di}, Attila and {Bapat}, Shreyas and {Barentsen}, Geert and {Baz{\'a}n}, Juanjo and {Biswas}, Manish and {Boquien}, M{\'e}d{\'e}ric and {Burke}, D.~J. and {Cara}, Daria and {Cara}, Mihai and {Conroy}, Kyle E. and {Conseil}, Simon and {Craig}, Matthew W. and {Cross}, Robert M. and {Cruz}, Kelle L. and {D'Eugenio}, Francesco and {Dencheva}, Nadia and {Devillepoix}, Hadrien A.~R. and {Dietrich}, J{\"o}rg P. and {Eigenbrot}, Arthur Davis and {Erben}, Thomas and {Ferreira}, Leonardo and {Foreman-Mackey}, Daniel and {Fox}, Ryan and {Freij}, Nabil and {Garg}, Suyog and {Geda}, Robel and {Glattly}, Lauren and {Gondhalekar}, Yash and {Gordon}, Karl D. and {Grant}, David and {Greenfield}, Perry and {Groener}, Austen M. and {Guest}, Steve and {Gurovich}, Sebastian and {Handberg}, Rasmus and {Hart}, Akeem and {Hatfield-Dodds}, Zac and {Homeier}, Derek and {Hosseinzadeh}, Griffin and {Jenness}, Tim and {Jones}, Craig K. and {Joseph}, Prajwel and {Kalmbach}, J. Bryce and {Karamehmetoglu}, Emir and {Ka{\l}uszy{\'n}ski}, Miko{\l}aj and {Kelley}, Michael S.~P. and {Kern}, Nicholas and {Kerzendorf}, Wolfgang E. and {Koch}, Eric W. and {Kulumani}, Shankar and {Lee}, Antony and {Ly}, Chun and {Ma}, Zhiyuan and {MacBride}, Conor and {Maljaars}, Jakob M. and {Muna}, Demitri and {Murphy}, N.~A. and {Norman}, Henrik and {O'Steen}, Richard and {Oman}, Kyle A. and {Pacifici}, Camilla and {Pascual}, Sergio and {Pascual-Granado}, J. and {Patil}, Rohit R. and {Perren}, Gabriel I. and {Pickering}, Timothy E. and {Rastogi}, Tanuj and {Roulston}, Benjamin R. and {Ryan}, Daniel F. and {Rykoff}, Eli S. and {Sabater}, Jose and {Sakurikar}, Parikshit and {Salgado}, Jes{\'u}s and {Sanghi}, Aniket and {Saunders}, Nicholas and {Savchenko}, Volodymyr and {Schwardt}, Ludwig and {Seifert-Eckert}, Michael and {Shih}, Albert Y. and {Jain}, Anany Shrey and {Shukla}, Gyanendra and {Sick}, Jonathan and {Simpson}, Chris and {Singanamalla}, Sudheesh and {Singer}, Leo P. and {Singhal}, Jaladh and {Sinha}, Manodeep and {Sip{\H{o}}cz}, Brigitta M. and {Spitler}, Lee R. and {Stansby}, David and {Streicher}, Ole and {{\v{S}}umak}, Jani and {Swinbank}, John D. and {Taranu}, Dan S. and {Tewary}, Nikita and {Tremblay}, Grant R. and {Val-Borro}, Miguel de and {Van Kooten}, Samuel J. and {Vasovi{\'c}}, Zlatan and {Verma}, Shresth and {de Miranda Cardoso}, Jos{\'e} Vin{\'\i}cius and {Williams}, Peter K.~G. and {Wilson}, Tom J. and {Winkel}, Benjamin and {Wood-Vasey}, W.~M. and {Xue}, Rui and {Yoachim}, Peter and {Zhang}, Chen and {Zonca}, Andrea and {Astropy Project Contributors}},
        title = "{The Astropy Project: Sustaining and Growing a Community-oriented Open-source Project and the Latest Major Release (v5.0) of the Core Package}",
      journal = {\apj},
         year = 2022,
        month = aug,
       volume = {935},
       number = {2},
          eid = {167},
        pages = {167},
          doi = {10.3847/1538-4357/ac7c74},
archivePrefix = {arXiv},
       eprint = {2206.14220},
 primaryClass = {astro-ph.IM},
       adsurl = {https://ui.adsabs.harvard.edu/abs/2022ApJ...935..167A}
}

@software{photutils,
author       = {Larry Bradley and
                Brigitta Sip{\H o}cz and
                Thomas Robitaille and
                Erik Tollerud and
                Z\`e Vin{\'{\i}}cius and
                Christoph Deil and
                Kyle Barbary and
                Tom J Wilson and
                Ivo Busko and
                Hans Moritz G{\"u}nther and
                Mihai Cara and
                Simon Conseil and
                Azalee Bostroem and
                Michael Droettboom and
                E. M. Bray and
                Lars Andersen Bratholm and
                P. L. Lim and
                Geert Barentsen and
                Matt Craig and
                Sergio Pascual and
                Gabriel Perren and
                Johnny Greco and
                Axel Donath and
                Miguel de Val-Borro and
                Wolfgang Kerzendorf and
                Yoonsoo P. Bach and
                Benjamin Alan Weaver and
                Francesco D'Eugenio and
                Harrison Souchereau and
                Leonardo Ferreira},
title        = {astropy/photutils: 1.0.0},
month        = sep,
year         = 2020,
publisher    = {Zenodo},
version      = {1.0.0},
doi          = {10.5281/zenodo.4044744},
url          = {https://doi.org/10.5281/zenodo.4044744}
}

@ARTICLE{riffel21,
       author = {{Riffel}, Rog{\'e}rio and {Mallmann}, Nicolas D. and {Ilha}, Gabriele S. and {Storchi-Bergmann}, Thaisa and {Riffel}, Rogemar A. and {Rembold}, Sandro B. and {Bizyaev}, Dmitry and {do Nascimento}, Janaina C. and {Schimoia}, Jaderson S. and {da Costa}, Luiz N. and {Boardman}, Nicholas Fraser and {Boquien}, M{\'e}d{\'e}ric and {Couto}, Guilherme S.},
        title = "{Determining star formation rates in active galactic nuclei hosts via stellar population synthesis}",
      journal = {\mnras},
         year = 2021,
        month = mar,
       volume = {501},
       number = {3},
        pages = {4064-4079},
          doi = {10.1093/mnras/staa3907},
archivePrefix = {arXiv},
       eprint = {2012.08472},
 primaryClass = {astro-ph.GA},
       adsurl = {https://ui.adsabs.harvard.edu/abs/2021MNRAS.501.4064R}
}

@ARTICLE{asari07,
       author = {{Asari}, N.~V. and {Cid Fernandes}, R. and {Stasi{\'n}ska}, G. and {Torres-Papaqui}, J.~P. and {Mateus}, A. and {Sodr{\'e}}, L. and {Schoenell}, W. and {Gomes}, J.~M.},
        title = "{The history of star-forming galaxies in the Sloan Digital Sky Survey}",
      journal = {\mnras},
         year = 2007,
        month = oct,
       volume = {381},
       number = {1},
        pages = {263-279},
          doi = {10.1111/j.1365-2966.2007.12255.x},
archivePrefix = {arXiv},
       eprint = {0707.3578},
 primaryClass = {astro-ph},
       adsurl = {https://ui.adsabs.harvard.edu/abs/2007MNRAS.381..263A}
}

@INPROCEEDINGS{weygaert&platen11,
       author = {{van de Weygaert}, Rien and {Platen}, Erwin},
        title = "{Cosmic Voids: Structure, Dynamics and Galaxies}",
    booktitle = {International Journal of Modern Physics Conference Series},
         year = 2011,
       series = {International Journal of Modern Physics Conference Series},
       volume = {1},
        month = jan,
        pages = {41-66},
          doi = {10.1142/S2010194511000092},
archivePrefix = {arXiv},
       eprint = {0912.2997},
 primaryClass = {astro-ph.CO},
       adsurl = {https://ui.adsabs.harvard.edu/abs/2011IJMPS...1...41V}
}

@ARTICLE{colless03,
       author = {{Colless}, Matthew and {Peterson}, Bruce A. and {Jackson}, Carole and {Peacock}, John A. and {Cole}, Shaun and {Norberg}, Peder and {Baldry}, Ivan K. and {Baugh}, Carlton M. and {Bland-Hawthorn}, Joss and {Bridges}, Terry and {Cannon}, Russell and {Collins}, Chris and {Couch}, Warrick and {Cross}, Nicholas and {Dalton}, Gavin and {De Propris}, Roberto and {Driver}, Simon P. and {Efstathiou}, George and {Ellis}, Richard S. and {Frenk}, Carlos S. and {Glazebrook}, Karl and {Lahav}, Ofer and {Lewis}, Ian and {Lumsden}, Stuart and {Maddox}, Steve and {Madgwick}, Darren and {Sutherland}, Will and {Taylor}, Keith},
        title = "{The 2dF Galaxy Redshift Survey: Final Data Release}",
      journal = {arXiv e-prints},
         year = 2003,
        month = jun,
          eid = {astro-ph/0306581},
        pages = {astro-ph/0306581},
          doi = {10.48550/arXiv.astro-ph/0306581},
archivePrefix = {arXiv},
       eprint = {astro-ph/0306581},
 primaryClass = {astro-ph},
       adsurl = {https://ui.adsabs.harvard.edu/abs/2003astro.ph..6581C}
}

@ARTICLE{tegmark04,
       author = {{Tegmark}, Max and {Blanton}, Michael R. and {Strauss}, Michael A. and {Hoyle}, Fiona and {Schlegel}, David and {Scoccimarro}, Roman and {Vogeley}, Michael S. and {Weinberg}, David H. and {Zehavi}, Idit and {Berlind}, Andreas and {Budavari}, Tam{\'a}s and {Connolly}, Andrew and {Eisenstein}, Daniel J. and {Finkbeiner}, Douglas and {Frieman}, Joshua A. and {Gunn}, James E. and {Hamilton}, Andrew J.~S. and {Hui}, Lam and {Jain}, Bhuvnesh and {Johnston}, David and {Kent}, Stephen and {Lin}, Huan and {Nakajima}, Reiko and {Nichol}, Robert C. and {Ostriker}, Jeremiah P. and {Pope}, Adrian and {Scranton}, Ryan and {Seljak}, Uro{\v{s}} and {Sheth}, Ravi K. and {Stebbins}, Albert and {Szalay}, Alexander S. and {Szapudi}, Istv{\'a}n and {Verde}, Licia and {Xu}, Yongzhong and {Annis}, James and {Bahcall}, Neta A. and {Brinkmann}, J. and {Burles}, Scott and {Castander}, Francisco J. and {Csabai}, Istvan and {Loveday}, Jon and {Doi}, Mamoru and {Fukugita}, Masataka and {Gott}, J. Richard, III and {Hennessy}, Greg and {Hogg}, David W. and {Ivezi{\'c}}, {\v{Z}}eljko and {Knapp}, Gillian R. and {Lamb}, Don Q. and {Lee}, Brian C. and {Lupton}, Robert H. and {McKay}, Timothy A. and {Kunszt}, Peter and {Munn}, Jeffrey A. and {O'Connell}, Liam and {Peoples}, John and {Pier}, Jeffrey R. and {Richmond}, Michael and {Rockosi}, Constance and {Schneider}, Donald P. and {Stoughton}, Christopher and {Tucker}, Douglas L. and {Vanden Berk}, Daniel E. and {Yanny}, Brian and {York}, Donald G. and {SDSS Collaboration}},
        title = "{The Three-Dimensional Power Spectrum of Galaxies from the Sloan Digital Sky Survey}",
      journal = {\apj},
         year = 2004,
        month = may,
       volume = {606},
       number = {2},
        pages = {702-740},
          doi = {10.1086/382125},
archivePrefix = {arXiv},
       eprint = {astro-ph/0310725},
 primaryClass = {astro-ph},
       adsurl = {https://ui.adsabs.harvard.edu/abs/2004ApJ...606..702T}
}

@ARTICLE{dubinski93,
       author = {{Dubinski}, John and {da Costa}, L.~N. and {Goldwirth}, D.~S. and {Lecar}, M. and {Piran}, T.},
        title = "{Void Evolution and the Large-Scale Structure}",
      journal = {\apj},
         year = 1993,
        month = jun,
       volume = {410},
        pages = {458},
          doi = {10.1086/172762},
       adsurl = {https://ui.adsabs.harvard.edu/abs/1993ApJ...410..458D}
}

@ARTICLE{weygaert&kampen93,
       author = {{van de Weygaert}, R. and {van Kampen}, E.},
        title = "{Voids in Gravitational Instability Scenarios - Part One - Global Density and Velocity Fields in an Einstein - De-Sitter Universe}",
      journal = {\mnras},
         year = 1993,
        month = jul,
       volume = {263},
        pages = {481},
          doi = {10.1093/mnras/263.2.481},
       adsurl = {https://ui.adsabs.harvard.edu/abs/1993MNRAS.263..481V}
}

@ARTICLE{colberg05,
       author = {{Colberg}, J{\"o}rg M. and {Sheth}, Ravi K. and {Diaferio}, Antonaldo and {Gao}, Liang and {Yoshida}, Naoki},
        title = "{Voids in a {\ensuremath{\Lambda}}CDM universe}",
      journal = {\mnras},
         year = 2005,
        month = jun,
       volume = {360},
       number = {1},
        pages = {216-226},
          doi = {10.1111/j.1365-2966.2005.09064.x},
archivePrefix = {arXiv},
       eprint = {astro-ph/0409162},
 primaryClass = {astro-ph},
       adsurl = {https://ui.adsabs.harvard.edu/abs/2005MNRAS.360..216C}
}

@ARTICLE{hoyle02,
       author = {{Hoyle}, Fiona and {Vogeley}, Michael S.},
        title = "{Voids in the Point Source Catalogue Survey and the Updated Zwicky Catalog}",
      journal = {\apj},
         year = 2002,
        month = feb,
       volume = {566},
       number = {2},
        pages = {641-651},
          doi = {10.1086/338340},
archivePrefix = {arXiv},
       eprint = {astro-ph/0109357},
 primaryClass = {astro-ph},
       adsurl = {https://ui.adsabs.harvard.edu/abs/2002ApJ...566..641H}
}

@ARTICLE{plionis02,
       author = {{Plionis}, Manolis and {Basilakos}, Spyros},
        title = "{The size and shape of local voids}",
      journal = {\mnras},
         year = 2002,
        month = feb,
       volume = {330},
       number = {2},
        pages = {399-404},
          doi = {10.1046/j.1365-8711.2002.05069.x},
archivePrefix = {arXiv},
       eprint = {astro-ph/0106491},
 primaryClass = {astro-ph},
       adsurl = {https://ui.adsabs.harvard.edu/abs/2002MNRAS.330..399P}
}

@ARTICLE{conroy05,
       author = {{Conroy}, Charlie and {Coil}, Alison L. and {White}, Martin and {Newman}, Jeffrey A. and {Yan}, Renbin and {Cooper}, Michael C. and {Gerke}, Brian F. and {Davis}, Marc and {Koo}, David C.},
        title = "{The DEEP2 Galaxy Redshift Survey: The Evolution of Void Statistics from z \raisebox{-0.5ex}\textasciitilde 1 to z \raisebox{-0.5ex}\textasciitilde 0}",
      journal = {\apj},
         year = 2005,
        month = dec,
       volume = {635},
       number = {2},
        pages = {990-1005},
          doi = {10.1086/497682},
archivePrefix = {arXiv},
       eprint = {astro-ph/0508250},
 primaryClass = {astro-ph},
       adsurl = {https://ui.adsabs.harvard.edu/abs/2005ApJ...635..990C}
}

@ARTICLE{szomoru96,
       author = {{Szomoru}, Arpad and {van Gorkom}, J.~H. and {Gregg}, Michael D. and {Strauss}, Michael A.},
        title = "{An HI Survey of the Bootes Void. II. The Analysis}",
      journal = {\aj},
         year = 1996,
        month = jun,
       volume = {111},
        pages = {2150},
          doi = {10.1086/117951},
archivePrefix = {arXiv},
       eprint = {astro-ph/9511128},
 primaryClass = {astro-ph},
       adsurl = {https://ui.adsabs.harvard.edu/abs/1996AJ....111.2150S}
}

@ARTICLE{el-ad97,
       author = {{El-Ad}, Hagai and {Piran}, Tsvi},
        title = "{Voids in the Large-Scale Structure}",
      journal = {\apj},
         year = 1997,
        month = dec,
       volume = {491},
       number = {2},
        pages = {421-435},
          doi = {10.1086/304973},
archivePrefix = {arXiv},
       eprint = {astro-ph/9702135},
 primaryClass = {astro-ph},
       adsurl = {https://ui.adsabs.harvard.edu/abs/1997ApJ...491..421E}
}

@ARTICLE{hoyle&vogeley04,
       author = {{Hoyle}, Fiona and {Vogeley}, Michael S.},
        title = "{Voids in the Two-Degree Field Galaxy Redshift Survey}",
      journal = {\apj},
         year = 2004,
        month = jun,
       volume = {607},
       number = {2},
        pages = {751-764},
          doi = {10.1086/386279},
archivePrefix = {arXiv},
       eprint = {astro-ph/0312533},
 primaryClass = {astro-ph},
       adsurl = {https://ui.adsabs.harvard.edu/abs/2004ApJ...607..751H}
}

@ARTICLE{sheth04,
       author = {{Sheth}, Ravi K. and {van de Weygaert}, Rien},
        title = "{A hierarchy of voids: much ado about nothing}",
      journal = {\mnras},
         year = 2004,
        month = may,
       volume = {350},
       number = {2},
        pages = {517-538},
          doi = {10.1111/j.1365-2966.2004.07661.x},
archivePrefix = {arXiv},
       eprint = {astro-ph/0311260},
 primaryClass = {astro-ph},
       adsurl = {https://ui.adsabs.harvard.edu/abs/2004MNRAS.350..517S}
}

@ARTICLE{kreckel12,
       author = {{Kreckel}, K. and {Platen}, E. and {Arag{\'o}n-Calvo}, M.~A. and {van Gorkom}, J.~H. and {van de Weygaert}, R. and {van der Hulst}, J.~M. and {Beygu}, B.},
        title = "{The Void Galaxy Survey: Optical Properties and H I Morphology and Kinematics}",
      journal = {\aj},
         year = 2012,
        month = jul,
       volume = {144},
       number = {1},
          eid = {16},
        pages = {16},
          doi = {10.1088/0004-6256/144/1/16},
archivePrefix = {arXiv},
       eprint = {1204.5185},
 primaryClass = {astro-ph.CO},
       adsurl = {https://ui.adsabs.harvard.edu/abs/2012AJ....144...16K}
}

@ARTICLE{alpaslan14,
       author = {{Alpaslan}, M. and {Robotham}, A.~S.~G. and {Obreschkow}, D. and {Penny}, S. and {Driver}, S. and {Norberg}, P. and {Brough}, S. and {Brown}, M. and {Cluver}, M. and {Holwerda}, B. and {Hopkins}, A.~M. and {van Kampen}, E. and {Kelvin}, L.~S. and {Lara-Lopez}, M.~A. and {Liske}, J. and {Loveday}, J. and {Mahajan}, S. and {Pimbblet}, K.},
        title = "{Galaxy and Mass Assembly (GAMA): fine filaments of galaxies detected within voids.}",
      journal = {\mnras},
         year = 2014,
        month = may,
       volume = {440},
        pages = {L106-L110},
          doi = {10.1093/mnrasl/slu019},
archivePrefix = {arXiv},
       eprint = {1401.7331},
 primaryClass = {astro-ph.CO},
       adsurl = {https://ui.adsabs.harvard.edu/abs/2014MNRAS.440L.106A}
}

@ARTICLE{goldberg&vogeley04,
       author = {{Goldberg}, David M. and {Vogeley}, Michael S.},
        title = "{Simulating Voids}",
      journal = {\apj},
         year = 2004,
        month = apr,
       volume = {605},
       number = {1},
        pages = {1-6},
          doi = {10.1086/382143},
archivePrefix = {arXiv},
       eprint = {astro-ph/0307191},
 primaryClass = {astro-ph},
       adsurl = {https://ui.adsabs.harvard.edu/abs/2004ApJ...605....1G}
}

@ARTICLE{plank20,
       author = {{Planck Collaboration} and {Aghanim}, N. and {Akrami}, Y. and {Arroja}, F. and {Ashdown}, M. and {Aumont}, J. and {Baccigalupi}, C. and {Ballardini}, M. and {Banday}, A.~J. and {Barreiro}, R.~B. and {Bartolo}, N. and {Basak}, S. and {Battye}, R. and {Benabed}, K. and {Bernard}, J. -P. and {Bersanelli}, M. and {Bielewicz}, P. and {Bock}, J.~J. and {Bond}, J.~R. and {Borrill}, J. and {Bouchet}, F.~R. and {Boulanger}, F. and {Bucher}, M. and {Burigana}, C. and {Butler}, R.~C. and {Calabrese}, E. and {Cardoso}, J. -F. and {Carron}, J. and {Casaponsa}, B. and {Challinor}, A. and {Chiang}, H.~C. and {Colombo}, L.~P.~L. and {Combet}, C. and {Contreras}, D. and {Crill}, B.~P. and {Cuttaia}, F. and {de Bernardis}, P. and {de Zotti}, G. and {Delabrouille}, J. and {Delouis}, J. -M. and {D{\'e}sert}, F. -X. and {Di Valentino}, E. and {Dickinson}, C. and {Diego}, J.~M. and {Donzelli}, S. and {Dor{\'e}}, O. and {Douspis}, M. and {Ducout}, A. and {Dupac}, X. and {Efstathiou}, G. and {Elsner}, F. and {En{\ss}lin}, T.~A. and {Eriksen}, H.~K. and {Falgarone}, E. and {Fantaye}, Y. and {Fergusson}, J. and {Fernandez-Cobos}, R. and {Finelli}, F. and {Forastieri}, F. and {Frailis}, M. and {Franceschi}, E. and {Frolov}, A. and {Galeotta}, S. and {Galli}, S. and {Ganga}, K. and {G{\'e}nova-Santos}, R.~T. and {Gerbino}, M. and {Ghosh}, T. and {Gonz{\'a}lez-Nuevo}, J. and {G{\'o}rski}, K.~M. and {Gratton}, S. and {Gruppuso}, A. and {Gudmundsson}, J.~E. and {Hamann}, J. and {Handley}, W. and {Hansen}, F.~K. and {Helou}, G. and {Herranz}, D. and {Hildebrandt}, S.~R. and {Hivon}, E. and {Huang}, Z. and {Jaffe}, A.~H. and {Jones}, W.~C. and {Karakci}, A. and {Keih{\"a}nen}, E. and {Keskitalo}, R. and {Kiiveri}, K. and {Kim}, J. and {Kisner}, T.~S. and {Knox}, L. and {Krachmalnicoff}, N. and {Kunz}, M. and {Kurki-Suonio}, H. and {Lagache}, G. and {Lamarre}, J. -M. and {Langer}, M. and {Lasenby}, A. and {Lattanzi}, M. and {Lawrence}, C.~R. and {Le Jeune}, M. and {Leahy}, J.~P. and {Lesgourgues}, J. and {Levrier}, F. and {Lewis}, A. and {Liguori}, M. and {Lilje}, P.~B. and {Lilley}, M. and {Lindholm}, V. and {L{\'o}pez-Caniego}, M. and {Lubin}, P.~M. and {Ma}, Y. -Z. and {Mac{\'\i}as-P{\'e}rez}, J.~F. and {Maggio}, G. and {Maino}, D. and {Mandolesi}, N. and {Mangilli}, A. and {Marcos-Caballero}, A. and {Maris}, M. and {Martin}, P.~G. and {Martinelli}, M. and {Mart{\'\i}nez-Gonz{\'a}lez}, E. and {Matarrese}, S. and {Mauri}, N. and {McEwen}, J.~D. and {Meerburg}, P.~D. and {Meinhold}, P.~R. and {Melchiorri}, A. and {Mennella}, A. and {Migliaccio}, M. and {Millea}, M. and {Mitra}, S. and {Miville-Desch{\^e}nes}, M. -A. and {Molinari}, D. and {Moneti}, A. and {Montier}, L. and {Morgante}, G. and {Moss}, A. and {Mottet}, S. and {M{\"u}nchmeyer}, M. and {Natoli}, P. and {N{\o}rgaard-Nielsen}, H.~U. and {Oxborrow}, C.~A. and {Pagano}, L. and {Paoletti}, D. and {Partridge}, B. and {Patanchon}, G. and {Pearson}, T.~J. and {Peel}, M. and {Peiris}, H.~V. and {Perrotta}, F. and {Pettorino}, V. and {Piacentini}, F. and {Polastri}, L. and {Polenta}, G. and {Puget}, J. -L. and {Rachen}, J.~P. and {Reinecke}, M. and {Remazeilles}, M. and {Renault}, C. and {Renzi}, A. and {Rocha}, G. and {Rosset}, C. and {Roudier}, G. and {Rubi{\~n}o-Mart{\'\i}n}, J.~A. and {Ruiz-Granados}, B. and {Salvati}, L. and {Sandri}, M. and {Savelainen}, M. and {Scott}, D. and {Shellard}, E.~P.~S. and {Shiraishi}, M. and {Sirignano}, C. and {Sirri}, G. and {Spencer}, L.~D. and {Sunyaev}, R. and {Suur-Uski}, A. -S. and {Tauber}, J.~A. and {Tavagnacco}, D. and {Tenti}, M. and {Terenzi}, L. and {Toffolatti}, L. and {Tomasi}, M. and {Trombetti}, T. and {Valiviita}, J. and {Van Tent}, B. and {Vibert}, L. and {Vielva}, P. and {Villa}, F. and {Vittorio}, N. and {Wandelt}, B.~D. and {Wehus}, I.~K. and {White}, M. and {White}, S.~D.~M. and {Zacchei}, A. and {Zonca}, A.},
        title = "{Planck 2018 results. I. Overview and the cosmological legacy of Planck}",
      journal = {\aap},
         year = 2020,
        month = sep,
       volume = {641},
          eid = {A1},
        pages = {A1},
          doi = {10.1051/0004-6361/201833880},
archivePrefix = {arXiv},
       eprint = {1807.06205},
 primaryClass = {astro-ph.CO},
       adsurl = {https://ui.adsabs.harvard.edu/abs/2020A&A...641A...1P}
}

@ARTICLE{bothun92,
       author = {{Bothun}, Gregory D. and {Geller}, Margaret J. and {Kurtz}, Michael J. and {Huchra}, John P. and {Schild}, Rudolph E.},
        title = "{The Velocity-Distance Relation for Galaxies on a Bubble}",
      journal = {\apj},
         year = 1992,
        month = aug,
       volume = {395},
        pages = {347},
          doi = {10.1086/171657},
       adsurl = {https://ui.adsabs.harvard.edu/abs/1992ApJ...395..347B}
}

@PROCEEDINGS{weygaert16,
        title = "{The Zeldovich Universe: Genesis and Growth of the Cosmic Web}",
    booktitle = {The Zeldovich Universe: Genesis and Growth of the Cosmic Web},
         year = 2016,
       editor = {{van de Weygaert}, R. and {Shandarin}, S. and {Saar}, E. and {Einasto}, J.},
       series = {IAU Symposium},
       volume = {308},
        month = oct,
          doi = {10.1017/S174392131601098X},
       adsurl = {https://ui.adsabs.harvard.edu/abs/2016IAUS..308.....V}
}

@ARTICLE{valles-perez21,
       author = {{Vall{\'e}s-P{\'e}rez}, David and {Quilis}, Vicent and {Planelles}, Susana},
        title = "{Void Replenishment: How Voids Accrete Matter Over Cosmic History}",
      journal = {\apjl},
         year = 2021,
        month = oct,
       volume = {920},
       number = {1},
          eid = {L2},
        pages = {L2},
          doi = {10.3847/2041-8213/ac2816},
archivePrefix = {arXiv},
       eprint = {2109.08165},
 primaryClass = {astro-ph.CO},
       adsurl = {https://ui.adsabs.harvard.edu/abs/2021ApJ...920L...2V}
}

@ARTICLE{bermejo24,
       author = {{Bermejo}, Raul and {Wilding}, Georg and {van de Weygaert}, Rien and {Jones}, Bernard J.~T. and {Vegter}, Gert and {Efstathiou}, Konstantinos},
        title = "{Topological bias: how haloes trace structural patterns in the cosmic web}",
      journal = {\mnras},
         year = 2024,
        month = apr,
       volume = {529},
       number = {4},
        pages = {4325-4353},
          doi = {10.1093/mnras/stae543},
archivePrefix = {arXiv},
       eprint = {2206.14655},
 primaryClass = {astro-ph.CO},
       adsurl = {https://ui.adsabs.harvard.edu/abs/2024MNRAS.529.4325B}
}

@ARTICLE{peebles01,
       author = {{Peebles}, P.~J.~E.},
        title = "{The Void Phenomenon}",
      journal = {\apj},
         year = 2001,
        month = aug,
       volume = {557},
       number = {2},
        pages = {495-504},
          doi = {10.1086/322254},
archivePrefix = {arXiv},
       eprint = {astro-ph/0101127},
 primaryClass = {astro-ph},
       adsurl = {https://ui.adsabs.harvard.edu/abs/2001ApJ...557..495P}
}

@ARTICLE{rojas04,
       author = {{Rojas}, Randall R. and {Vogeley}, Michael S. and {Hoyle}, Fiona and {Brinkmann}, Jon},
        title = "{Photometric Properties of Void Galaxies in the Sloan Digital Sky Survey}",
      journal = {\apj},
         year = 2004,
        month = dec,
       volume = {617},
       number = {1},
        pages = {50-63},
          doi = {10.1086/425225},
archivePrefix = {arXiv},
       eprint = {astro-ph/0307274},
 primaryClass = {astro-ph},
       adsurl = {https://ui.adsabs.harvard.edu/abs/2004ApJ...617...50R}
}

@ARTICLE{croton05,
       author = {{Croton}, Darren J. and {Farrar}, Glennys R. and {Norberg}, Peder and {Colless}, Matthew and {Peacock}, John A. and {Baldry}, I.~K. and {Baugh}, C.~M. and {Bland-Hawthorn}, J. and {Bridges}, T. and {Cannon}, R. and {Cole}, S. and {Collins}, C. and {Couch}, W. and {Dalton}, G. and {De Propris}, R. and {Driver}, S.~P. and {Efstathiou}, G. and {Ellis}, R.~S. and {Frenk}, C.~S. and {Glazebrook}, K. and {Jackson}, C. and {Lahav}, O. and {Lewis}, I. and {Lumsden}, S. and {Maddox}, S. and {Madgwick}, D. and {Peterson}, B.~A. and {Sutherland}, W. and {Taylor}, K.},
        title = "{The 2dF Galaxy Redshift Survey: luminosity functions by density environment and galaxy type}",
      journal = {\mnras},
         year = 2005,
        month = jan,
       volume = {356},
       number = {3},
        pages = {1155-1167},
          doi = {10.1111/j.1365-2966.2004.08546.x},
archivePrefix = {arXiv},
       eprint = {astro-ph/0407537},
 primaryClass = {astro-ph},
       adsurl = {https://ui.adsabs.harvard.edu/abs/2005MNRAS.356.1155C}
}

@ARTICLE{hoyle12,
       author = {{Hoyle}, Fiona and {Vogeley}, M.~S. and {Pan}, D.},
        title = "{Photometric properties of void galaxies in the Sloan Digital Sky Survey Data Release 7}",
      journal = {\mnras},
         year = 2012,
        month = nov,
       volume = {426},
       number = {4},
        pages = {3041-3050},
          doi = {10.1111/j.1365-2966.2012.21943.x},
archivePrefix = {arXiv},
       eprint = {1205.1843},
 primaryClass = {astro-ph.CO},
       adsurl = {https://ui.adsabs.harvard.edu/abs/2012MNRAS.426.3041H}
}

@ARTICLE{beygu16,
       author = {{Beygu}, B. and {Kreckel}, K. and {van der Hulst}, J.~M. and {Jarrett}, T.~H. and {Peletier}, R. and {van de Weygaert}, R. and {van Gorkom}, J.~H. and {Aragon-Calvo}, M.~A.},
        title = "{The void galaxy survey: Star formation properties}",
      journal = {\mnras},
         year = 2016,
        month = may,
       volume = {458},
       number = {1},
        pages = {394-409},
          doi = {10.1093/mnras/stw280},
archivePrefix = {arXiv},
       eprint = {1601.08228},
 primaryClass = {astro-ph.GA},
       adsurl = {https://ui.adsabs.harvard.edu/abs/2016MNRAS.458..394B}
}

@ARTICLE{douglass18,
       author = {{Douglass}, Kelly A. and {Vogeley}, Michael S. and {Cen}, Renyue},
        title = "{Influence of the Void Environment on Chemical Abundances in Dwarf Galaxies and Implications for Connecting Star Formation and Halo Mass}",
      journal = {\apj},
         year = 2018,
        month = sep,
       volume = {864},
       number = {2},
          eid = {144},
        pages = {144},
          doi = {10.3847/1538-4357/aad86e},
archivePrefix = {arXiv},
       eprint = {1706.07099},
 primaryClass = {astro-ph.GA},
       adsurl = {https://ui.adsabs.harvard.edu/abs/2018ApJ...864..144D}
}

@ARTICLE{florez21,
       author = {{Florez}, Jonathan and {Berlind}, Andreas A. and {Kannappan}, Sheila J. and {Stark}, David V. and {Eckert}, Kathleen D. and {Calderon}, Victor F. and {Moffett}, Amanda J. and {Campbell}, Duncan and {Sinha}, Manodeep},
        title = "{Void Galaxies Follow a Distinct Evolutionary Path in the Environmental COntext Catalog}",
      journal = {\apj},
         year = 2021,
        month = jan,
       volume = {906},
       number = {2},
          eid = {97},
        pages = {97},
          doi = {10.3847/1538-4357/abca9f},
archivePrefix = {arXiv},
       eprint = {2011.08276},
 primaryClass = {astro-ph.GA},
       adsurl = {https://ui.adsabs.harvard.edu/abs/2021ApJ...906...97F}
}

@ARTICLE{pandey21,
       author = {{Pandey}, Divya and {Saha}, Kanak and {Pradhan}, Ananta C.},
        title = "{The Ultraviolet Deep Imaging Survey of Galaxies in the Bootes Void. I. Catalog, Color-Magnitude Relations, and Star Formation}",
      journal = {\apj},
         year = 2021,
        month = oct,
       volume = {919},
       number = {2},
          eid = {101},
        pages = {101},
          doi = {10.3847/1538-4357/ac1078},
archivePrefix = {arXiv},
       eprint = {2107.01774},
 primaryClass = {astro-ph.GA},
       adsurl = {https://ui.adsabs.harvard.edu/abs/2021ApJ...919..101P}
}

@ARTICLE{medrano22,
       author = {{Rodr{\'\i}guez Medrano}, Agust{\'\i}n M. and {Paz}, Dante J. and {Stasyszyn}, Federico A. and {Ruiz}, Andr{\'e}s N.},
        title = "{Imprints of the cosmic void evolution on the baryon processes inside galaxy haloes}",
      journal = {\mnras},
         year = 2022,
        month = apr,
       volume = {511},
       number = {2},
        pages = {2688-2701},
          doi = {10.1093/mnras/stac127},
archivePrefix = {arXiv},
       eprint = {2109.13378},
 primaryClass = {astro-ph.CO},
       adsurl = {https://ui.adsabs.harvard.edu/abs/2022MNRAS.511.2688R}
}

@ARTICLE{rosas-guevara22,
       author = {{Rosas-Guevara}, Yetli and {Tissera}, Patricia and {Lagos}, Claudia del P. and {Paillas}, Enrique and {Padilla}, Nelson},
        title = "{Revealing the properties of void galaxies and their assembly using the EAGLE simulation}",
      journal = {\mnras},
         year = 2022,
        month = nov,
       volume = {517},
       number = {1},
        pages = {712-731},
          doi = {10.1093/mnras/stac2583},
archivePrefix = {arXiv},
       eprint = {2204.04565},
 primaryClass = {astro-ph.GA},
       adsurl = {https://ui.adsabs.harvard.edu/abs/2022MNRAS.517..712R}
}

@ARTICLE{patiri06,
       author = {{Patiri}, Santiago G. and {Prada}, Francisco and {Holtzman}, Jon and {Klypin}, Anatoly and {Betancort-Rijo}, Juan},
        title = "{The properties of galaxies in voids}",
      journal = {\mnras},
         year = 2006,
        month = nov,
       volume = {372},
       number = {4},
        pages = {1710-1720},
          doi = {10.1111/j.1365-2966.2006.10975.x},
archivePrefix = {arXiv},
       eprint = {astro-ph/0605703},
 primaryClass = {astro-ph},
       adsurl = {https://ui.adsabs.harvard.edu/abs/2006MNRAS.372.1710P}
}

@ARTICLE{moorman14,
       author = {{Moorman}, Crystal M. and {Vogeley}, Michael S. and {Hoyle}, Fiona and {Pan}, Danny C. and {Haynes}, Martha P. and {Giovanelli}, Riccardo},
        title = "{The H I mass function and velocity width function of void galaxies in the Arecibo Legacy Fast ALFA Survey}",
      journal = {\mnras},
         year = 2014,
        month = nov,
       volume = {444},
       number = {4},
        pages = {3559-3570},
          doi = {10.1093/mnras/stu1674},
archivePrefix = {arXiv},
       eprint = {1408.3392},
 primaryClass = {astro-ph.GA},
       adsurl = {https://ui.adsabs.harvard.edu/abs/2014MNRAS.444.3559M}
}

@ARTICLE{liu15,
       author = {{Liu}, Chen-Xu and {Pan}, Danny C. and {Hao}, Lei and {Hoyle}, Fiona and {Constantin}, Anca and {Vogeley}, Michael S.},
        title = "{Spectral Properties of Galaxies in Void Regions}",
      journal = {\apj},
         year = 2015,
        month = sep,
       volume = {810},
       number = {2},
          eid = {165},
        pages = {165},
          doi = {10.1088/0004-637X/810/2/165},
archivePrefix = {arXiv},
       eprint = {1509.04430},
 primaryClass = {astro-ph.GA},
       adsurl = {https://ui.adsabs.harvard.edu/abs/2015ApJ...810..165L}
}

@ARTICLE{douglass&vogeley17,
       author = {{Douglass}, Kelly A. and {Vogeley}, Michael S.},
        title = "{Determining the Large-scale Environmental Dependence of Gas-phase Metallicity in Dwarf Galaxies}",
      journal = {\apj},
         year = 2017,
        month = jan,
       volume = {834},
       number = {2},
          eid = {186},
        pages = {186},
          doi = {10.3847/1538-4357/834/2/186},
archivePrefix = {arXiv},
       eprint = {1604.08599},
 primaryClass = {astro-ph.GA},
       adsurl = {https://ui.adsabs.harvard.edu/abs/2017ApJ...834..186D}
}

@ARTICLE{douglass19,
       author = {{Douglass}, Kelly A. and {Smith}, Jacob A. and {Demina}, Regina},
        title = "{The Influence of the Void Environment on the Ratio of Dark Matter Halo Mass to Stellar Mass in SDSS MaNGA Galaxies}",
      journal = {\apj},
         year = 2019,
        month = dec,
       volume = {886},
       number = {2},
          eid = {153},
        pages = {153},
          doi = {10.3847/1538-4357/ab4bce},
archivePrefix = {arXiv},
       eprint = {1906.08327},
 primaryClass = {astro-ph.GA},
       adsurl = {https://ui.adsabs.harvard.edu/abs/2019ApJ...886..153D}
}

@ARTICLE{wegner19,
       author = {{Wegner}, Gary A. and {Salzer}, John J. and {Taylor}, Joanna M. and {Hirschauer}, Alec S.},
        title = "{Metal Abundances and Star Formation Rates of Emission-line Galaxies in and around the Bo{\"o}tes Void}",
      journal = {\apj},
         year = 2019,
        month = sep,
       volume = {883},
       number = {1},
          eid = {29},
        pages = {29},
          doi = {10.3847/1538-4357/ab3a3c},
archivePrefix = {arXiv},
       eprint = {1908.07539},
 primaryClass = {astro-ph.GA},
       adsurl = {https://ui.adsabs.harvard.edu/abs/2019ApJ...883...29W}
}

@ARTICLE{dominguez-gomez22,
       author = {{Dom{\'\i}nguez-G{\'o}mez}, J. and {Lisenfeld}, U. and {P{\'e}rez}, I. and {L{\'o}pez-S{\'a}nchez}, {\'A}. R. and {Duarte Puertas}, S. and {Falc{\'o}n-Barroso}, J. and {Kreckel}, K. and {Peletier}, R.~F. and {Ruiz-Lara}, T. and {van de Weygaert}, R. and {van der Hulst}, J.~M. and {Verley}, S.},
        title = "{CO-CAVITY pilot survey: Molecular gas and star formation in void galaxies}",
      journal = {\aap},
         year = 2022,
        month = feb,
       volume = {658},
          eid = {A124},
        pages = {A124},
          doi = {10.1051/0004-6361/202141888},
archivePrefix = {arXiv},
       eprint = {2111.06844},
 primaryClass = {astro-ph.GA},
       adsurl = {https://ui.adsabs.harvard.edu/abs/2022A&A...658A.124D}
}

@ARTICLE{curtis24,
       author = {{Curtis}, Olivia and {McDonough}, Bryanne and {Brainerd}, Tereasa G.},
        title = "{Properties of Voids and Void Galaxies in the TNG300 Simulation}",
      journal = {\apj},
         year = 2024,
        month = feb,
       volume = {962},
       number = {1},
          eid = {58},
        pages = {58},
          doi = {10.3847/1538-4357/ad18b4},
archivePrefix = {arXiv},
       eprint = {2401.02322},
 primaryClass = {astro-ph.GA},
       adsurl = {https://ui.adsabs.harvard.edu/abs/2024ApJ...962...58C}
}

@ARTICLE{argudo-fernandez15,
       author = {{Argudo-Fern{\'a}ndez}, M. and {Verley}, S. and {Bergond}, G. and {Duarte Puertas}, S. and {Ramos Carmona}, E. and {Sabater}, J. and {Fern{\'a}ndez Lorenzo}, M. and {Espada}, D. and {Sulentic}, J. and {Ruiz}, J.~E. and {Leon}, S.},
        title = "{Catalogues of isolated galaxies, isolated pairs, and isolated triplets in the local Universe}",
      journal = {\aap},
         year = 2015,
        month = jun,
       volume = {578},
          eid = {A110},
        pages = {A110},
          doi = {10.1051/0004-6361/201526016},
archivePrefix = {arXiv},
       eprint = {1504.00117},
 primaryClass = {astro-ph.GA},
       adsurl = {https://ui.adsabs.harvard.edu/abs/2015A&A...578A.110A}
}

@ARTICLE{beygu13,
       author = {{Beygu}, B. and {Kreckel}, K. and {van de Weygaert}, R. and {van der Hulst}, J.~M. and {van Gorkom}, J.~H.},
        title = "{An Interacting Galaxy System along a Filament in a Void}",
      journal = {\aj},
         year = 2013,
        month = may,
       volume = {145},
       number = {5},
          eid = {120},
        pages = {120},
          doi = {10.1088/0004-6256/145/5/120},
archivePrefix = {arXiv},
       eprint = {1303.0538},
 primaryClass = {astro-ph.CO},
       adsurl = {https://ui.adsabs.harvard.edu/abs/2013AJ....145..120B}
}

@ARTICLE{sutter12,
       author = {{Sutter}, P.~M. and {Lavaux}, Guilhem and {Wandelt}, Benjamin D. and {Weinberg}, David H.},
        title = "{A Public Void Catalog from the SDSS DR7 Galaxy Redshift Surveys Based on the Watershed Transform}",
      journal = {\apj},
         year = 2012,
        month = dec,
       volume = {761},
       number = {1},
          eid = {44},
        pages = {44},
          doi = {10.1088/0004-637X/761/1/44},
archivePrefix = {arXiv},
       eprint = {1207.2524},
 primaryClass = {astro-ph.CO},
       adsurl = {https://ui.adsabs.harvard.edu/abs/2012ApJ...761...44S}
}

@ARTICLE{karachentseva79,
       author = {{Karachentseva}, V.~E. and {Karachentsev}, I.~D. and {Shcherbanovsky}, A.~L.},
        title = "{Isolated triplet of galaxies.}",
      journal = {Astrofizicheskie Issledovaniia Izvestiya Spetsial'noj Astrofizicheskoj Observatorii},
         year = 1979,
        month = jan,
       volume = {11},
        pages = {3-39},
       adsurl = {https://ui.adsabs.harvard.edu/abs/1979AISAO..11....3K}
}

@INPROCEEDINGS{karachentseva&karachentseva00,
       author = {{Karachentseva}, V.~E. and {Karachentsev}, I.~D.},
        title = "{Southern triplets of galaxies}",
    booktitle = {IAU Colloq. 174: Small Galaxy Groups},
         year = 2000,
       editor = {{Valtonen}, Mauri J. and {Flynn}, Chris},
       series = {Astronomical Society of the Pacific Conference Series},
       volume = {209},
        month = jan,
        pages = {11},
       adsurl = {https://ui.adsabs.harvard.edu/abs/2000ASPC..209...11K}
}

@ARTICLE{elyiv09,
       author = {{Elyiv}, A. and {Melnyk}, O. and {Vavilova}, I.},
        title = "{High-order 3D Voronoi tessellation for identifying isolated galaxies, pairs and triplets}",
      journal = {\mnras},
         year = 2009,
        month = apr,
       volume = {394},
       number = {3},
        pages = {1409-1418},
          doi = {10.1111/j.1365-2966.2008.14150.x},
archivePrefix = {arXiv},
       eprint = {0810.5100},
 primaryClass = {astro-ph},
       adsurl = {https://ui.adsabs.harvard.edu/abs/2009MNRAS.394.1409E}
}

@ARTICLE{makarov&karachentsev09,
       author = {{Makarov}, D.~I. and {Karachentsev}, I.~D.},
        title = "{Galaxy triplets in the local supercluster}",
      journal = {Astrophysical Bulletin},
         year = 2009,
        month = jan,
       volume = {64},
       number = {1},
        pages = {24-49},
          doi = {10.1134/S1990341309010027},
archivePrefix = {arXiv},
       eprint = {0908.1357},
 primaryClass = {astro-ph.CO},
       adsurl = {https://ui.adsabs.harvard.edu/abs/2009AstBu..64...24M}
}

@ARTICLE{chernin00,
       author = {{Chernin}, A.~D. and {Dolgachev}, V.~P. and {Domozhilova}, L.~M.},
        title = "{Wide triplets of galaxies: collapse on spatial scales of \raisebox{-0.5ex}\textasciitilde1Mpc}",
      journal = {\mnras},
         year = 2000,
        month = dec,
       volume = {319},
       number = {3},
        pages = {851-859},
          doi = {10.1046/j.1365-8711.2000.03909.x},
       adsurl = {https://ui.adsabs.harvard.edu/abs/2000MNRAS.319..851C}
}

@ARTICLE{hernandez-toledo11,
       author = {{Hern{\'a}ndez-Toledo}, H.~M. and {M{\'e}ndez-Hern{\'a}ndez}, H. and {Aceves}, H. and {Olgu{\'\i}n}, L.},
        title = "{BVRI Surface Photometry of Isolated Galaxy Triplets}",
      journal = {\aj},
         year = 2011,
        month = mar,
       volume = {141},
       number = {3},
          eid = {74},
        pages = {74},
          doi = {10.1088/0004-6256/141/3/74},
       adsurl = {https://ui.adsabs.harvard.edu/abs/2011AJ....141...74H}
}

@ARTICLE{duplancic13,
       author = {{Duplancic}, Fernanda and {O'Mill}, Ana Laura and {Lambas}, Diego G. and {Sodr{\'e}}, Laerte and {Alonso}, Sol},
        title = "{Galaxy triplets in Sloan Digital Sky Survey Data Release 7 - II. A connection with compact groups?}",
      journal = {\mnras},
         year = 2013,
        month = aug,
       volume = {433},
       number = {4},
        pages = {3547-3558},
          doi = {10.1093/mnras/stt985},
archivePrefix = {arXiv},
       eprint = {1306.0891},
 primaryClass = {astro-ph.CO},
       adsurl = {https://ui.adsabs.harvard.edu/abs/2013MNRAS.433.3547D}
}

@ARTICLE{feng16,
       author = {{Feng}, Shuai and {Shao}, Zheng-Yi and {Shen}, Shi-Yin and {Argudo-Fern{\'a}ndez}, Maria and {Wu}, Hong and {Lam}, Man-I. and {Yang}, Ming and {Yuan}, Fang-Ting},
        title = "{An isolated compact galaxy triplet}",
      journal = {Research in Astronomy and Astrophysics},
         year = 2016,
        month = may,
       volume = {16},
       number = {5},
          eid = {72},
        pages = {72},
          doi = {10.1088/1674-4527/16/5/072},
archivePrefix = {arXiv},
       eprint = {1512.02439},
 primaryClass = {astro-ph.GA},
       adsurl = {https://ui.adsabs.harvard.edu/abs/2016RAA....16...72F}
}

@ARTICLE{duplancic15,
       author = {{Duplancic}, Fernanda and {Alonso}, Sol and {Lambas}, Diego G. and {O'Mill}, Ana Laura},
        title = "{Galaxy triplets in Sloan Digital Sky Survey Data Release 7 - III. Analysis of configuration and dynamics}",
      journal = {\mnras},
         year = 2015,
        month = feb,
       volume = {447},
       number = {2},
        pages = {1399-1406},
          doi = {10.1093/mnras/stu2518},
archivePrefix = {arXiv},
       eprint = {1412.0022},
 primaryClass = {astro-ph.GA},
       adsurl = {https://ui.adsabs.harvard.edu/abs/2015MNRAS.447.1399D}
}

@ARTICLE{costa-duarte16,
       author = {{Costa-Duarte}, M.~V. and {O'Mill}, A.~L. and {Duplancic}, F. and {Sodr{\'e}}, L. and {Lambas}, D.~G.},
        title = "{Dissecting galaxy triplets in the Sloan Digital Sky Survey Data Release 10 - I. Stellar populations and emission line analysis}",
      journal = {\mnras},
         year = 2016,
        month = jul,
       volume = {459},
       number = {3},
        pages = {2539-2549},
          doi = {10.1093/mnras/stw816},
archivePrefix = {arXiv},
       eprint = {1604.02149},
 primaryClass = {astro-ph.GA},
       adsurl = {https://ui.adsabs.harvard.edu/abs/2016MNRAS.459.2539C}
}

@ARTICLE{trofimov&chernin95,
       author = {{Trofimov}, A.~V. and {Chernin}, A.~D.},
        title = "{Wide triplets of galaxies and the problem of hidden mass.}",
      journal = {\azh},
         year = 1995,
        month = jun,
       volume = {72},
        pages = {308-317},
       adsurl = {https://ui.adsabs.harvard.edu/abs/1995AZh....72..308T}
}

@ARTICLE{vasquez-bustos23,
       author = {{V{\'a}squez-Bustos}, P. and {Argudo-Fernandez}, M. and {Grajales-Medina}, D. and {Duarte Puertas}, S. and {Verley}, S.},
        title = "{Understanding the role of morphology and environment in the dynamical evolution of isolated galaxy triplets}",
      journal = {\aap},
         year = 2023,
        month = feb,
       volume = {670},
          eid = {A63},
        pages = {A63},
          doi = {10.1051/0004-6361/202245297},
archivePrefix = {arXiv},
       eprint = {2211.10290},
 primaryClass = {astro-ph.GA},
       adsurl = {https://ui.adsabs.harvard.edu/abs/2023A&A...670A..63V}
}

@ARTICLE{planck19,
       author = {{Kozmanyan}, Arpine and {Bourdin}, Herv{\'e} and {Mazzotta}, Pasquale and
         {Rasia}, Elena and {Sereno}, Mauro},
        title = "{Deriving the Hubble constant using Planck and XMM-Newton observations of galaxy clusters}",
      journal = {\aap},
         year = 2019,
        month = jan,
       volume = {621},
          eid = {A34},
        pages = {A34},
          doi = {10.1051/0004-6361/201833879},
archivePrefix = {arXiv},
       eprint = {1809.09560},
 primaryClass = {astro-ph.CO},
       adsurl = {https://ui.adsabs.harvard.edu/abs/2019A&A...621A..34K}
}

@ARTICLE{inla,
       author = {{Gonz{\'a}lez-Gait{\'a}n}, S. and {de Souza}, R.~S. and {Krone-Martins}, A. and {Cameron}, E. and {Coelho}, P. and {Galbany}, L. and {Ishida}, E.~E.~O. and {COIN Collaboration}},
        title = "{Spatial field reconstruction with INLA: application to IFU galaxy data}",
      journal = {\mnras},
         year = 2019,
        month = jan,
       volume = {482},
       number = {3},
        pages = {3880-3891},
          doi = {10.1093/mnras/sty2881},
archivePrefix = {arXiv},
       eprint = {1802.06280},
 primaryClass = {astro-ph.IM},
       adsurl = {https://ui.adsabs.harvard.edu/abs/2019MNRAS.482.3880G}
}

@ARTICLE{rojas05,
       author = {{Rojas}, Randall R. and {Vogeley}, Michael S. and {Hoyle}, Fiona and {Brinkmann}, Jon},
        title = "{Spectroscopic Properties of Void Galaxies in the Sloan Digital Sky Survey}",
      journal = {\apj},
         year = 2005,
        month = may,
       volume = {624},
       number = {2},
        pages = {571-585},
          doi = {10.1086/428476},
archivePrefix = {arXiv},
       eprint = {astro-ph/0409074},
 primaryClass = {astro-ph},
       adsurl = {https://ui.adsabs.harvard.edu/abs/2005ApJ...624..571R}
}

@ARTICLE{chambers16,
       author = {{Chambers}, K.~C. and {Magnier}, E.~A. and {Metcalfe}, N. and {Flewelling}, H.~A. and {Huber}, M.~E. and {Waters}, C.~Z. and {Denneau}, L. and {Draper}, P.~W. and {Farrow}, D. and {Finkbeiner}, D.~P. and {Holmberg}, C. and {Koppenhoefer}, J. and {Price}, P.~A. and {Rest}, A. and {Saglia}, R.~P. and {Schlafly}, E.~F. and {Smartt}, S.~J. and {Sweeney}, W. and {Wainscoat}, R.~J. and {Burgett}, W.~S. and {Chastel}, S. and {Grav}, T. and {Heasley}, J.~N. and {Hodapp}, K.~W. and {Jedicke}, R. and {Kaiser}, N. and {Kudritzki}, R. -P. and {Luppino}, G.~A. and {Lupton}, R.~H. and {Monet}, D.~G. and {Morgan}, J.~S. and {Onaka}, P.~M. and {Shiao}, B. and {Stubbs}, C.~W. and {Tonry}, J.~L. and {White}, R. and {Ba{\~n}ados}, E. and {Bell}, E.~F. and {Bender}, R. and {Bernard}, E.~J. and {Boegner}, M. and {Boffi}, F. and {Botticella}, M.~T. and {Calamida}, A. and {Casertano}, S. and {Chen}, W. -P. and {Chen}, X. and {Cole}, S. and {Deacon}, N. and {Frenk}, C. and {Fitzsimmons}, A. and {Gezari}, S. and {Gibbs}, V. and {Goessl}, C. and {Goggia}, T. and {Gourgue}, R. and {Goldman}, B. and {Grant}, P. and {Grebel}, E.~K. and {Hambly}, N.~C. and {Hasinger}, G. and {Heavens}, A.~F. and {Heckman}, T.~M. and {Henderson}, R. and {Henning}, T. and {Holman}, M. and {Hopp}, U. and {Ip}, W. -H. and {Isani}, S. and {Jackson}, M. and {Keyes}, C.~D. and {Koekemoer}, A.~M. and {Kotak}, R. and {Le}, D. and {Liska}, D. and {Long}, K.~S. and {Lucey}, J.~R. and {Liu}, M. and {Martin}, N.~F. and {Masci}, G. and {McLean}, B. and {Mindel}, E. and {Misra}, P. and {Morganson}, E. and {Murphy}, D.~N.~A. and {Obaika}, A. and {Narayan}, G. and {Nieto-Santisteban}, M.~A. and {Norberg}, P. and {Peacock}, J.~A. and {Pier}, E.~A. and {Postman}, M. and {Primak}, N. and {Rae}, C. and {Rai}, A. and {Riess}, A. and {Riffeser}, A. and {Rix}, H.~W. and {R{\"o}ser}, S. and {Russel}, R. and {Rutz}, L. and {Schilbach}, E. and {Schultz}, A.~S.~B. and {Scolnic}, D. and {Strolger}, L. and {Szalay}, A. and {Seitz}, S. and {Small}, E. and {Smith}, K.~W. and {Soderblom}, D.~R. and {Taylor}, P. and {Thomson}, R. and {Taylor}, A.~N. and {Thakar}, A.~R. and {Thiel}, J. and {Thilker}, D. and {Unger}, D. and {Urata}, Y. and {Valenti}, J. and {Wagner}, J. and {Walder}, T. and {Walter}, F. and {Watters}, S.~P. and {Werner}, S. and {Wood-Vasey}, W.~M. and {Wyse}, R.},
        title = "{The Pan-STARRS1 Surveys}",
      journal = {arXiv e-prints},
         year = 2016,
        month = dec,
          eid = {arXiv:1612.05560},
        pages = {arXiv:1612.05560},
          doi = {10.48550/arXiv.1612.05560},
archivePrefix = {arXiv},
       eprint = {1612.05560},
 primaryClass = {astro-ph.IM},
       adsurl = {https://ui.adsabs.harvard.edu/abs/2016arXiv161205560C}
}

@INPROCEEDINGS{weygaert11,
       author = {{van de Weygaert}, R. and {Kreckel}, K. and {Platen}, E. and {Beygu}, B. and {van Gorkom}, J.~H. and {van der Hulst}, J.~M. and {Arag{\'o}n-Calvo}, M.~A. and {Peebles}, P.~J.~E. and {Jarrett}, T. and {Rhee}, G. and {Kova{\v{c}}}, K. and {Yip}, C. -W.},
        title = "{The Void Galaxy Survey}",
    booktitle = {Environment and the Formation of Galaxies: 30 years later},
         year = 2011,
       editor = {{Ferreras}, Ignacio and {Pasquali}, Anna},
       series = {Astrophysics and Space Science Proceedings},
       volume = {27},
        month = jan,
        pages = {17},
          doi = {10.1007/978-3-642-20285-8_3},
archivePrefix = {arXiv},
       eprint = {1101.4187},
 primaryClass = {astro-ph.CO},
       adsurl = {https://ui.adsabs.harvard.edu/abs/2011ASSP...27...17V}
}

@ARTICLE{morfometryka,
       author = {{Ferrari}, F. and {de Carvalho}, R.~R. and {Trevisan}, M.},
        title = "{Morfometryka{\textemdash}A New Way of Establishing Morphological Classification of Galaxies}",
      journal = {\apj},
         year = 2015,
        month = nov,
       volume = {814},
       number = {1},
          eid = {55},
        pages = {55},
          doi = {10.1088/0004-637X/814/1/55},
archivePrefix = {arXiv},
       eprint = {1509.05430},
 primaryClass = {astro-ph.GA},
       adsurl = {https://ui.adsabs.harvard.edu/abs/2015ApJ...814...55F}
}

@ARTICLE{abraham94,
       author = {{Abraham}, Roberto G. and {Valdes}, Francisco and {Yee}, H.~K.~C. and {van den Bergh}, Sidney},
        title = "{The Morphologies of Distant Galaxies. I. an Automated Classification System}",
      journal = {\apj},
         year = 1994,
        month = sep,
       volume = {432},
        pages = {75},
          doi = {10.1086/174550},
       adsurl = {https://ui.adsabs.harvard.edu/abs/1994ApJ...432...75A}
}

@ARTICLE{abraham96,
       author = {{Abraham}, Roberto G. and {van den Bergh}, Sidney and {Glazebrook}, Karl and {Ellis}, Richard S. and {Santiago}, Basilio X. and {Surma}, Peter and {Griffiths}, Richard E.},
        title = "{The Morphologies of Distant Galaxies. II. Classifications from the Hubble Space Telescope Medium Deep Survey}",
      journal = {\apjs},
         year = 1996,
        month = nov,
       volume = {107},
        pages = {1},
          doi = {10.1086/192352},
       adsurl = {https://ui.adsabs.harvard.edu/abs/1996ApJS..107....1A}
}

@ARTICLE{conselice00,
       author = {{Conselice}, Christopher J. and {Bershady}, Matthew A. and {Jangren}, Anna},
        title = "{The Asymmetry of Galaxies: Physical Morphology for Nearby and High-Redshift Galaxies}",
      journal = {\apj},
         year = 2000,
        month = feb,
       volume = {529},
       number = {2},
        pages = {886-910},
          doi = {10.1086/308300},
archivePrefix = {arXiv},
       eprint = {astro-ph/9907399},
 primaryClass = {astro-ph},
       adsurl = {https://ui.adsabs.harvard.edu/abs/2000ApJ...529..886C}
}

@ARTICLE{lotz04,
       author = {{Lotz}, Jennifer M. and {Primack}, Joel and {Madau}, Piero},
        title = "{A New Nonparametric Approach to Galaxy Morphological Classification}",
      journal = {\aj},
         year = 2004,
        month = jul,
       volume = {128},
       number = {1},
        pages = {163-182},
          doi = {10.1086/421849},
archivePrefix = {arXiv},
       eprint = {astro-ph/0311352},
 primaryClass = {astro-ph},
       adsurl = {https://ui.adsabs.harvard.edu/abs/2004AJ....128..163L}
}

@BOOK{sersic68,
       author = {{Sersic}, Jose Luis},
        title = "{Atlas de Galaxias Australes}",
         year = 1968,
       adsurl = {https://ui.adsabs.harvard.edu/abs/1968adga.book.....S}
}

@ARTICLE{lucatelli19,
       author = {{Lucatelli}, Geferson and {Ferrari}, Fabricio},
        title = "{Galaxy structural analysis with the curvature of the brightness profile}",
      journal = {\mnras},
         year = 2019,
        month = oct,
       volume = {489},
       number = {1},
        pages = {1161-1180},
          doi = {10.1093/mnras/stz2154},
archivePrefix = {arXiv},
       eprint = {1907.10188},
 primaryClass = {astro-ph.GA},
       adsurl = {https://ui.adsabs.harvard.edu/abs/2019MNRAS.489.1161L}
}

@ARTICLE{mihos&hernquist94,
       author = {{Mihos}, J. Christopher and {Hernquist}, Lars},
        title = "{Triggering of Starbursts in Galaxies by Minor Mergers}",
      journal = {\apjl},
         year = 1994,
        month = apr,
       volume = {425},
        pages = {L13},
          doi = {10.1086/187299},
       adsurl = {https://ui.adsabs.harvard.edu/abs/1994ApJ...425L..13M}
}

@ARTICLE{mihos&hernquist96,
       author = {{Mihos}, J. Christopher and {Hernquist}, Lars},
        title = "{Gasdynamics and Starbursts in Major Mergers}",
      journal = {\apj},
         year = 1996,
        month = jun,
       volume = {464},
        pages = {641},
          doi = {10.1086/177353},
archivePrefix = {arXiv},
       eprint = {astro-ph/9512099},
 primaryClass = {astro-ph},
       adsurl = {https://ui.adsabs.harvard.edu/abs/1996ApJ...464..641M}
}

@INCOLLECTION{duc&renaud13,
       author = {{Duc}, Pierre-Alain and {Renaud}, Florent},
        title = "{Tides in Colliding Galaxies}",
    booktitle = {Lecture Notes in Physics, Berlin Springer Verlag},
         year = 2013,
       editor = {{Souchay}, Jean and {Mathis}, St{\'e}phane and {Tokieda}, Tadashi},
       volume = {861},
        pages = {327},
          doi = {10.1007/978-3-642-32961-6_9},
       adsurl = {https://ui.adsabs.harvard.edu/abs/2013LNP...861..327D}
}

@ARTICLE{pan12,
       author = {{Pan}, Danny C. and {Vogeley}, Michael S. and {Hoyle}, Fiona and {Choi}, Yun-Young and {Park}, Changbom},
        title = "{Cosmic voids in Sloan Digital Sky Survey Data Release 7}",
      journal = {\mnras},
         year = 2012,
        month = apr,
       volume = {421},
       number = {2},
        pages = {926-934},
          doi = {10.1111/j.1365-2966.2011.20197.x},
archivePrefix = {arXiv},
       eprint = {1103.4156},
 primaryClass = {astro-ph.CO},
       adsurl = {https://ui.adsabs.harvard.edu/abs/2012MNRAS.421..926P}
}

@ARTICLE{wechsler18,
       author = {{Wechsler}, Risa H. and {Tinker}, Jeremy L.},
        title = "{The Connection Between Galaxies and Their Dark Matter Halos}",
      journal = {\araa},
         year = 2018,
        month = sep,
       volume = {56},
        pages = {435-487},
          doi = {10.1146/annurev-astro-081817-051756},
archivePrefix = {arXiv},
       eprint = {1804.03097},
 primaryClass = {astro-ph.GA},
       adsurl = {https://ui.adsabs.harvard.edu/abs/2018ARA&A..56..435W}
}

\begin{appendix}

\section{Elliptical isophotes}

\begin{figure}
    \centering
    \includegraphics[width=\linewidth]{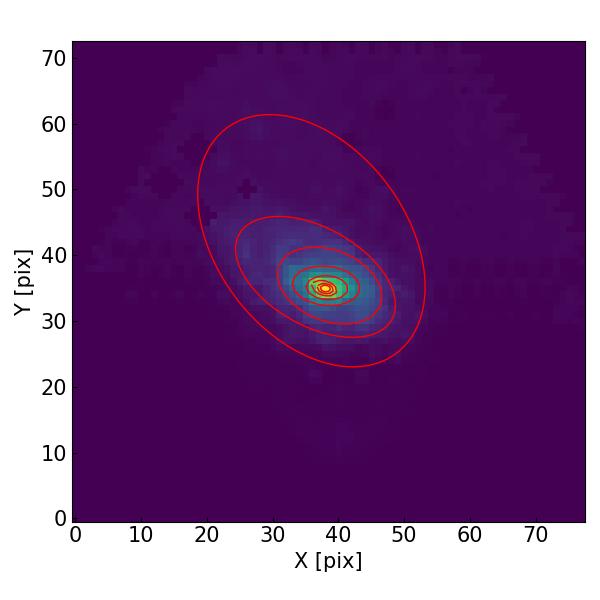}
    \caption{Map of integrated fluxes of the galaxy VGS31b. The red ellipses represent the isophotes fit with {\sc photutils}.}
    \label{elipses}
\end{figure}

All the galaxies in the triplets seem to have some sort of asymmetry. Still, we can fit elliptical isophotes using the Python library \href{https://photutils.readthedocs.io/en/stable/}{\sc photutils} \citep{photutils}, an \href{https://www.astropy.org/}{\sc astropy} package \citep{astropy} for photometry. This helps us to divide the galaxy into different regions (annuli) and analyse each one separately. We applied this methodology to the maps of the integrated fluxes of the spectra of the galaxies. Figure \ref{elipses} shows the example for the galaxy VGS31b.

By computing the mean ages and metallicities for the spaxels between two ellipses, we can obtain profiles for stellar ages and metallicities. Since the center, eccentricity, and angle of the isophotes can vary largely for the most asymmetric and clumpy galaxies. It means that the curves we compute are not exactly radial profiles, but rather measurements for regions with similar surface brightness. We choose to keep it this way because we verify that in almost all of the 6 galaxies, the ellipses change significantly as their SMA increases, and they present clear asymmetries.

\begin{figure}
    \centering
    \includegraphics[width=\linewidth]{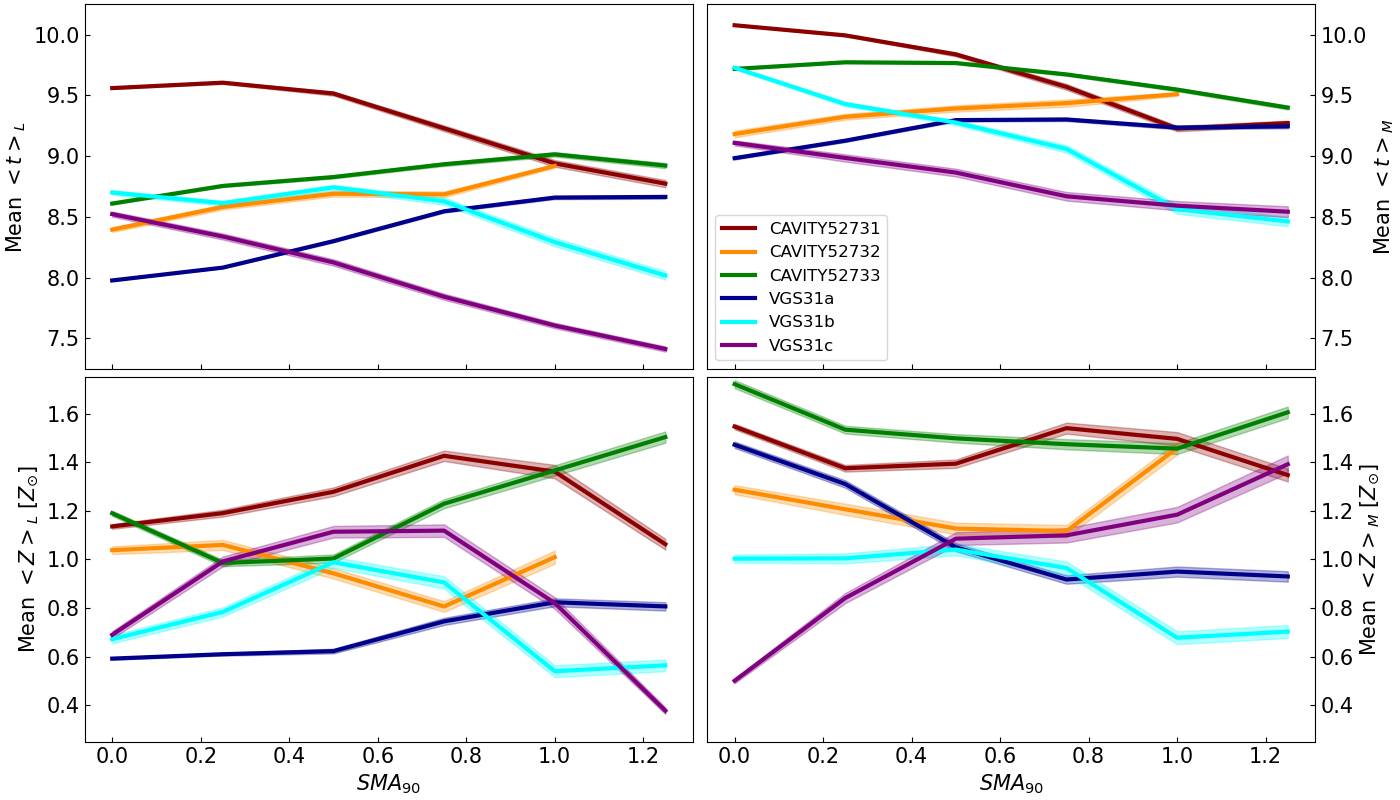}
    \caption{Mean radial profiles of $<\log t>_L$ (top left), $<\log t>_M$ (top right), $<Z>_L$ (bottom left), and $<Z>_M$ (bottom right). The curves were computed after performing a Monte Carlo, perturbing the mean value in each annulus 1000 times and then computing the mean of the 1000 generated curves. The shaded regions are the error of the mean.}
    \label{radial_profiles}
\end{figure}

Additionally to computing the mean ages or metallicities for different annuli, we perform a simple Monte Carlo method by perturbing the points 1000 times. For each point, we assume a Gaussian distribution with $\sigma$ equal to its respective standard error. In the end, we have the mean of the 1000 curves generated for each galaxy, along with the error of the mean. The SMAs are normalized by the SMA of the isophote that contains 90\% of the total light of the galaxy ($\text{SMA}_{90}$). Figure \ref{radial_profiles} presents the profiles for the individual galaxies in our sample, both light and mass-weighted.

There is no pattern for either the mean ages or the mean stellar metallicities profiles. Regarding age, half of the galaxies present an increasing profile and the other half a decreasing profile. The curves for $<\log t>_M$ are higher than the ones for $<\log t>_L$, as expected, since the mass-weighting highlights older stellar populations. In the case of CAVITY52733, the $<\log t>_L$ profile is increasing, while the $<\log t>_M$ starts decreasing for radii larger than $\sim 0.5R_{90}$. 2 galaxies present negative age profiles (VGS31b and VGS31c), which is not common. For instance, dwarf early-type galaxies of comparable mass in the Virgo cluster show age gradients always positive or nearly flat \citep{bidaran23}.

The mean $<Z>_L$ and $<Z>_M$ curves seem more complex. There are profiles that have a positive gradient in the inner parts of the galaxy and a negative in the outskirts. There are others that have the opposite behavior. Again, there is no clear pattern, but our sample is very small, and each galaxy may have passed through distinct evolutionary paths and had distinct SFHs.

\section{Age bin maps}

\begin{figure}
    \centering
    \includegraphics[width=\linewidth]{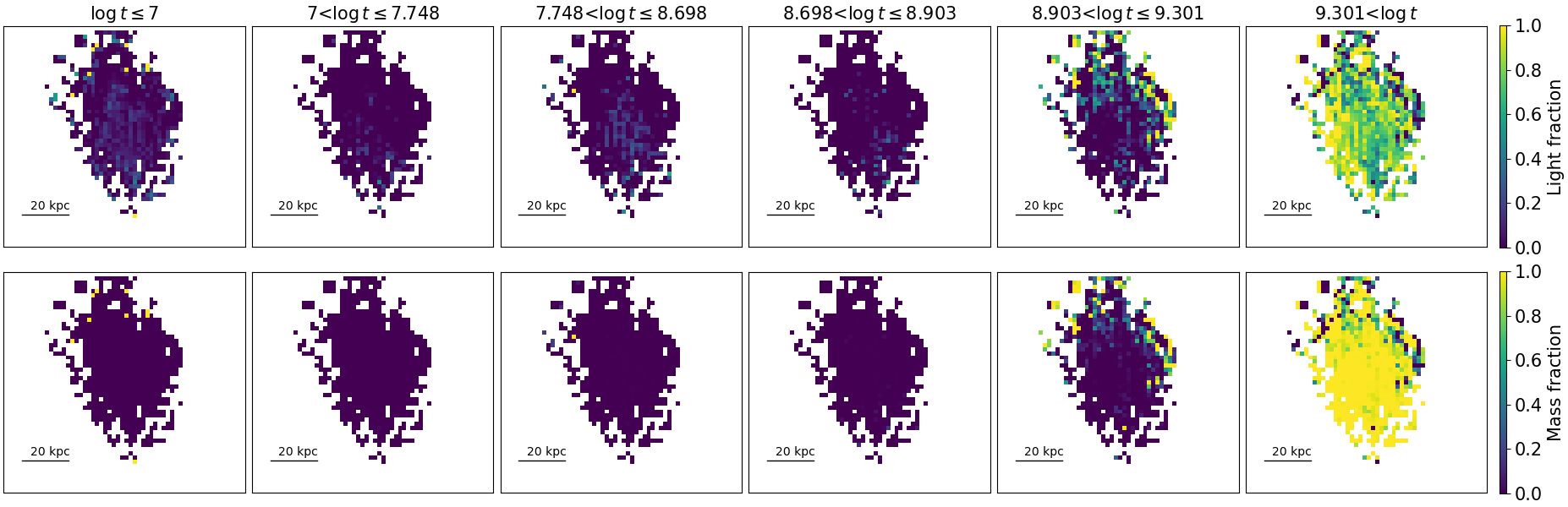}
    \includegraphics[width=\linewidth]{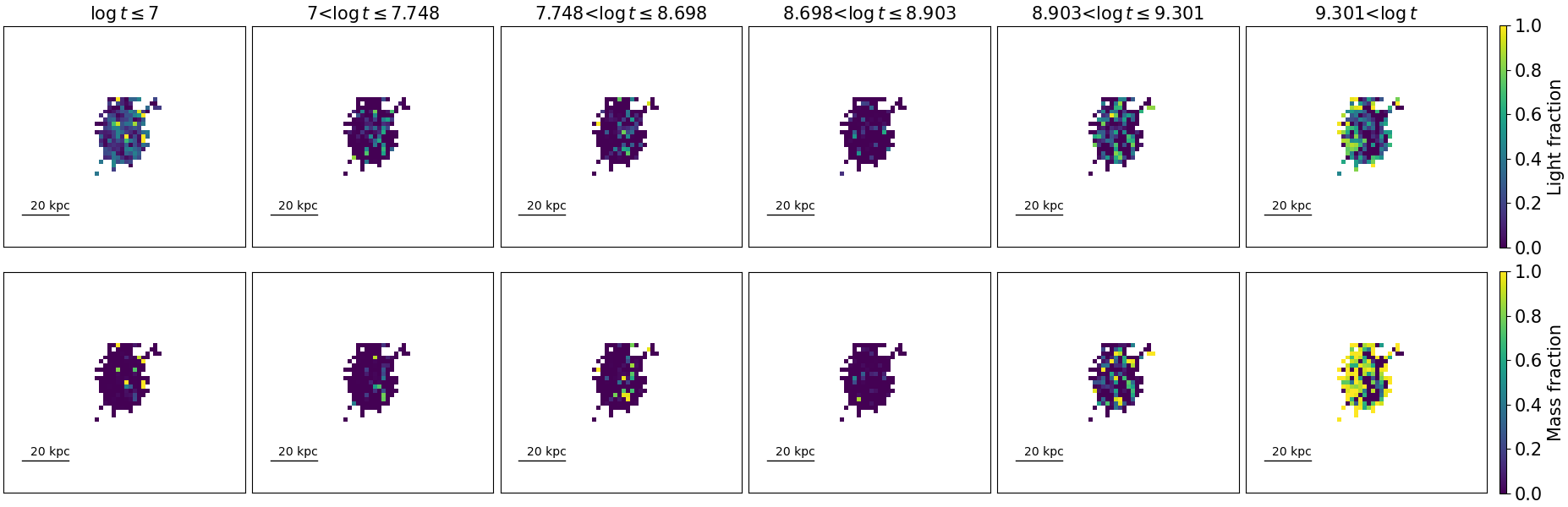}
    \includegraphics[width=\linewidth]{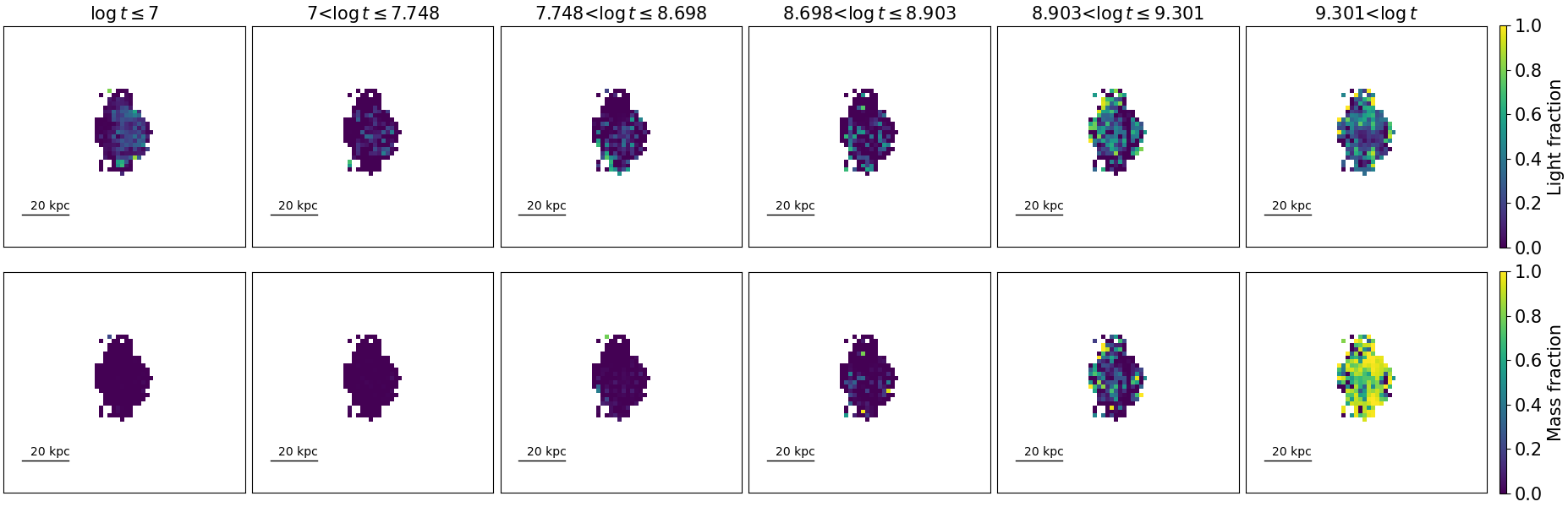}
    \caption{2D maps of light and mass fractions of stellar populations binned by age. Each panel is analogous to a bin in a histogram, so this is equivalent to a "spatially resolved histogram". The top galaxy is CAVITY52731, the middle one is CAVITY52732, and the bottom one is CAVITY52733.}
    \label{age_bins}
\end{figure}

\begin{figure}
    \centering
    \includegraphics[width=\linewidth]{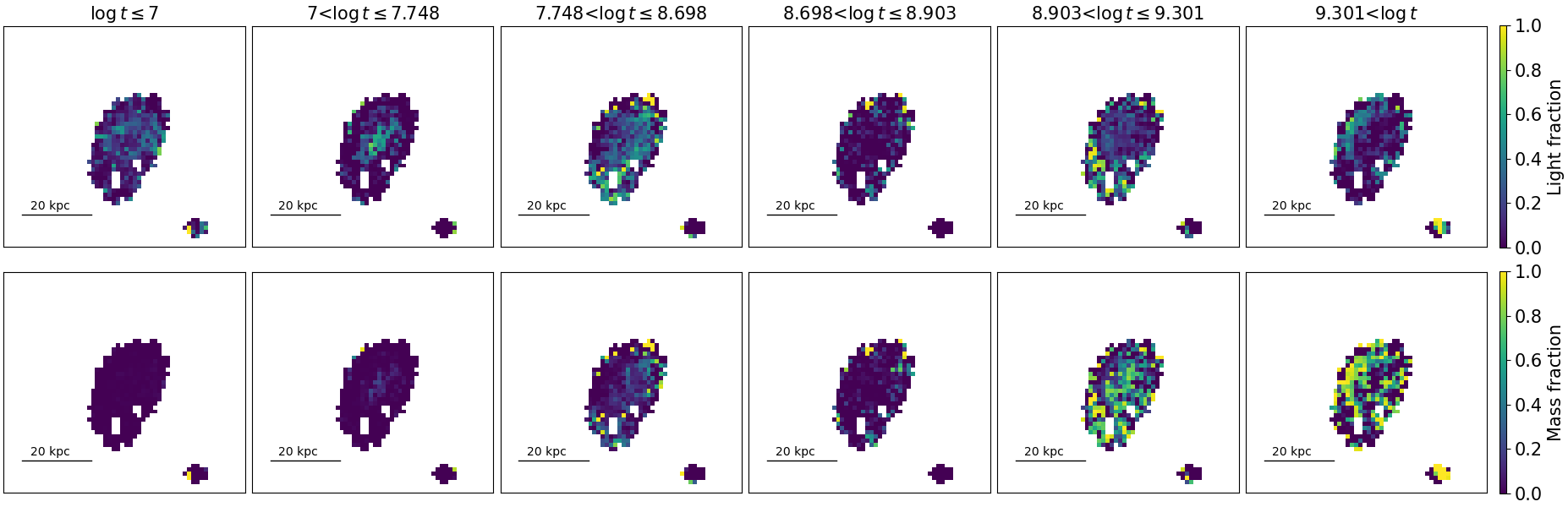}
    \includegraphics[width=\linewidth]{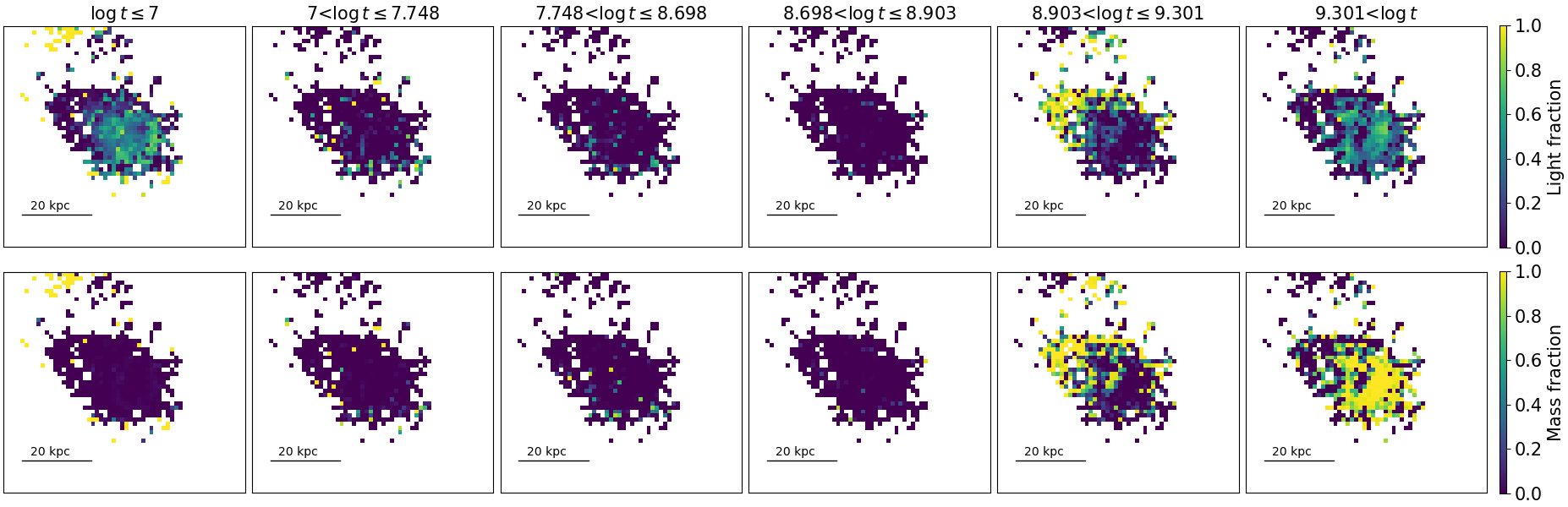}
    \includegraphics[width=\linewidth]{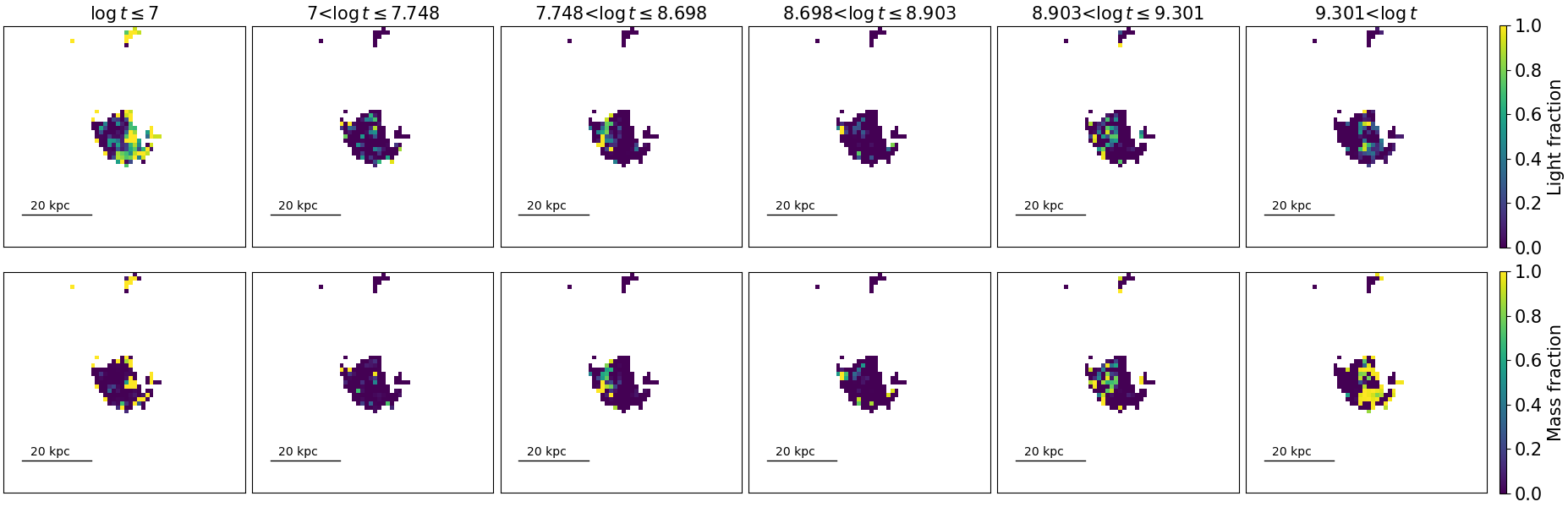}
    \caption{Same as Fig. \ref{age_bins} but for (from top to bottom) VGS31a, VSG31b, and VGS31c.}
    \label{age_bins_2}
\end{figure}

This section presents the age bins maps for the galaxies in our sample (Figs. \ref{age_bins} and \ref{age_bins_2}). By binning the stellar populations by age, disregarding metallicities, we obtain the SFHs of the galaxies in the form of the fraction of mass (or light) enclosed in each population. We present it in a different manner, however. Instead of curves or histograms, we show 2D maps with such fractions, with each panel in the image representing a bin of age.

We observe, for instance, most of the mass and light in CAVITY52731 is present in the population with $t > 2$ Gyr. For the rest of the galaxies, we see significant mass enclosed in population with $800 \,\text{Myr} < t \leq 2 \,\text{Gyr}$, which is in accordance with the behavior of the mass assembly functions. Additionally, we also see significant light fractions produced by young populations ($t < 100 \,\text{Myr}$) in the five star-forming galaxies, consistent with their nebular emission.

\section{Comparison with MaNGA triplets}

\begin{figure}
    \centering
    \includegraphics[width=0.55\linewidth]{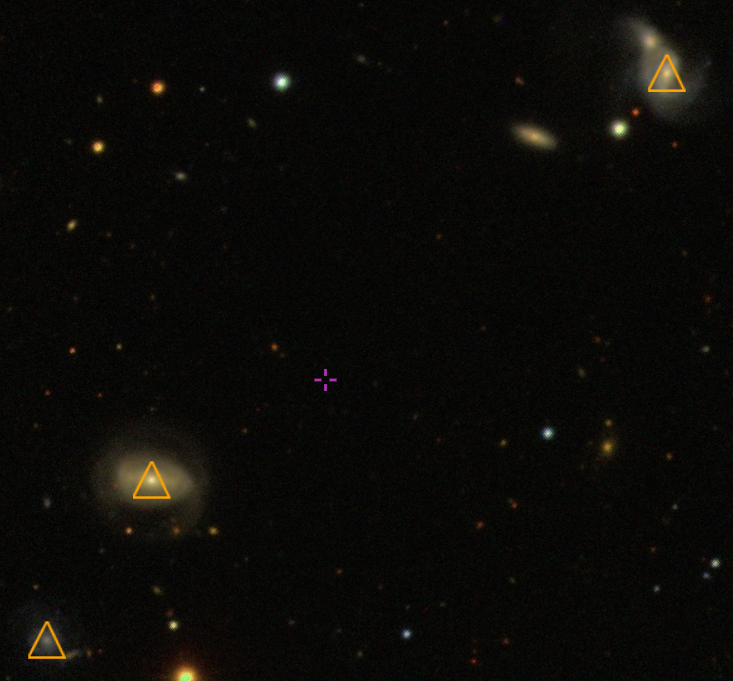} \includegraphics[width=0.42152\linewidth]{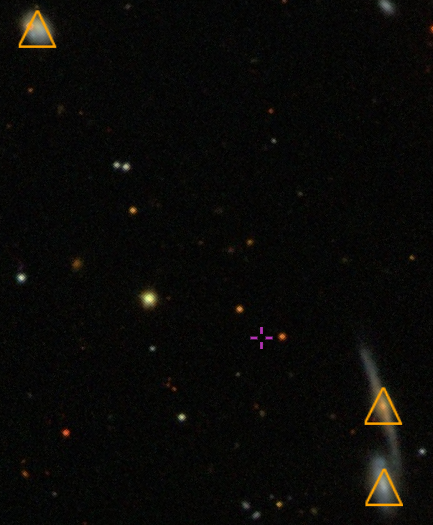}
    \caption{Three color images, taken from SDSS SkyViewer, of triplets from SIT catalog \citep{argudo-fernandez15}: SIT10 (left) and SIT178 (right).}
    \label{SIT}
\end{figure}

In this section we present a brief qualitative comparison between our triplets and two triplets from the SIT catalog \citep{argudo-fernandez15}. To the best of our knowledge, there is no sample of IFU observed triplets that could be used as a control sample to make direct comparisons. Among the 315 SIT galaxies, only two systems have all three galaxies observed by MaNGA. Those are SIT 10 and SIT 178 (see Fig. \ref{SIT}).

Both systems have a late-type central (most massive) galaxy. This is already distinct from our sample, as VSG31 has a late-type central galaxy, but CAVITY5273X have an early-type central galaxy. Additionally, their geometrical configurations are very distinct. Both SIT triplets consist of a pair of close galaxies in projection, with a third, further member. The systems seem less compact than VGS31 and CAVITY5273X.

In conclusion, the number of IFU observed triplets is still very low, making it difficult to get to statistical conclusions and compare specific cases with bigger samples. Further spatially resolved SPS should also be applied to these MaNGA triplets to enhance the quantity and quality of information we have about isolated triplets.

\end{appendix}

\end{document}